\documentclass[11pt, oneside, a4paper]{article}

\usepackage[left=2cm,right=2cm,top=2cm,bottom=2.4cm]{geometry}
\usepackage{amsmath, amsfonts}
\usepackage{mathtools}
\usepackage{bm, bbm}
\usepackage{setspace}
\usepackage{float}
\usepackage{enumitem}
\usepackage{authblk}   % Author affiliations
\usepackage{url}
\usepackage{chngcntr}

\usepackage{tikz}
\usetikzlibrary{positioning, calc, shapes}

\usepackage{rotating}
\usepackage{arydshln}
\newcounter{example}
\newenvironment{example}[1][]{\refstepcounter{example}\par\medskip
   \noindent \textit{Example~\theexample #1} \rmfamily}{\medskip}
  
\newcounter{algorithm}
\newenvironment{algorithm}[1][]{\refstepcounter{algorithm}\par\medskip
   \noindent \textbf{Algorithm~\thealgorithm #1} \rmfamily}{\medskip}

\newcommand\Tstrut{\rule{0pt}{2.6ex}}         % = top strut
\usepackage{multirow}

\usepackage{hyperref}
\usepackage{natbib}
\allowdisplaybreaks

\title{{Graphical approaches for the control of generalised error rates}}

\author[1]{D.\ S.\ Robertson}
\author[1,2]{J.\ M.\ S.\ Wason}
\author[3,4]{F.\ Bretz}
\affil[1]{\small MRC Biostatistics Unit, University of Cambridge, Cambridge, UK}
\affil[2]{\small Institute of Health and Society, Newcastle University, Newcastle, UK}
\affil[3]{\small Statistical Methodology, Novartis Pharma AG, CH-4002 Basel, Switzerland}
\affil[4]{\small Section for Medical Statistics, Medical University of Vienna, Vienna, Austria}

\date{}

\begin{document}

\maketitle

\begin{abstract}

\noindent When simultaneously testing multiple hypotheses, the usual approach in the context of confirmatory clinical trials is to control the familywise error rate (FWER), which bounds the probability of making at least one false rejection. In many trial settings, these hypotheses will additionally have a hierarchical structure that reflects the relative importance and links between different clinical objectives. The graphical approach of~\citet{Bretz2009} is a flexible and easily communicable way of controlling the FWER while respecting complex trial objectives and multiple structured hypotheses. However, the FWER can be a very stringent criteria that leads to procedures with a low power, and may not be appropriate in exploratory trial settings. This motivates controlling generalised error rates, particularly when the number of hypotheses tested is no longer very small. We consider the generalised familywise error rate ($k$-FWER), which is the probability of making~$k$ or more false rejections, as well as the tail probability of the false discovery proportion (FDP), which is the probability that the proportion of false rejections is greater than some threshold. We also consider asymptotic control of the false discovery rate (FDR), which is the expectation of the FDP. In this paper, we show how to control these generalised error rates when using the graphical approach and its extensions. We demonstrate the utility of the resulting graphical procedures on three clinical trial case studies. \\

\end{abstract}


\vspace{1.75cm}

\noindent Address correspondence to D.\ S.\ Robertson,  MRC Biostatistics Unit, University of Cambridge, IPH Forvie Site, Robinson Way, Cambridge CB2 0SR, UK; E-mail: david.robertson@mrc-bsu.cam.ac.uk

%%%%%%%%%%%%%%%%%%%%%%%%%%%%%%%%%%%%%%%%%%%%%%%%%%%

\section{Introduction}
\label{sec:intro}

%%% Multiple testing in clinical trials

In modern clinical trials, it is increasingly common to test multiple hypotheses simultaneously. This multiplicity is driven by evaluating multiple therapies in parallel, the identification of multiple subgroups and the measurement of multiple endpoints. Given that these multiple hypotheses are assessed simultaneously, there is a strong emphasis on controlling the total number or proportion of false positives (i.e.\ type~I errors) in some way. Indeed, for confirmatory clinical trials, regulatory guidelines state that the familywise error rate (FWER) should be strongly controlled~\citep{FDA2017, EMA2017}. This ensures that the maximum probability of making at least one type~I error is below some pre-specified level (under any configuration of the parameters being tested). %It is difficult to definitively demonstrate strong FWER control with simulations due to the dimensionality of the parameter space, and hence theoretical results are needed.

%%% Complex trial objectives and multiple structured hypotheses

The increase in multiplicity in clinical trials also tends to go hand-in-hand with an increase in the complexity of the objectives and structure of the hypotheses tested. A key setting where this occurs is when measuring multiple endpoints to answer distinct (but related) clinical questions. The corresponding hypotheses often fit naturally within a hierarchical structure that reflects the relevant importance and links between the clinical questions that the trial aims to answer. For example, in a trial with both a primary and secondary hypothesis, the trialist may only wish to test the secondary hypothesis if the primary hypothesis is first rejected. More complex hierarchical structures can be formed as the number of hypotheses increases.

%%% Introduce graphical approach
Many methods have been developed for FWER control that respect complex trial objectives and multiple structured hypotheses. A highly flexible framework for doing so is the graphical approach to hypothesis testing, as proposed independently by~\citet{Bretz2009} and~\citet{Burman2009}. In the framework of~\citet{Bretz2009}, vertices represent the null hypotheses and weights represent the local significance levels, which are propagated through weighted, directed edges. The resulting multiple testing procedure can be tailored to structured families of hypotheses with arbitrary dependence between the hypotheses, and allows the visualisation of complex decision strategies in an easily communicable way. Many well-known procedures for FWER control are special cases of the graphical approach, such as the fixed sequence (or hierarchical) test~\citep{Maurer1995}, the Holm procedure~\citep{Holm1979}, the Hochberg procedure~\citep{Xi2019} and several gatekeeping procedures~\citep{Dmitrienko2003, Dmitrienko2008, Li2019}.

%%% Introduce generalised error rates
% FWER is a stringent criterion

However, controlling the FWER is a very stringent criterion, especially as the number of hypotheses increases. By controlling the probability of even a single type~I error, the power of FWER-controlling procedures can be very low, with little chance of any of the individual hypotheses being rejected. While strong FWER control is appropriate in confirmatory contexts, in exploratory trial settings such strict criterion may not be necessary. Indeed, as reflected in the FDA (2017) guidance on multiple endpoints in clinical trials~\citep{FDA2017}, exploratory analyses can be included in a trial to explore and generate new hypotheses. Since such exploratory hypotheses will often be followed up with confirmatory testing, strict FWER control at the exploratory stage is no longer necessary.

\citet{Westfall2018} expand on this argument, by categorising the hypothesis tests in a typical clinical trial into families of ``efficacy'', ``safety'' and ``exploratory'' tests. For the efficacy family, the primary endpoints and main secondary endpoints are the basis of regulatory approval and labelling, and hence require strong FWER control. However, there may also be ``lesser interest tests" (for example, multiple time point analyses), where FWER controlling methods are not needed. Nonetheless, the authors note that some form of multiplicity adjustment would strengthen the claims made for this set of tests. For the safety family, serious and known treatment-related adverse events (AEs) do not require multiplicity adjustment (since type~II errors are of much greater concern). However, for all other AEs, the authors state that there is a clear need to recognise the multiplicity problem, and note that the use of the FDR may be more appropriate here. Finally, for the family of exploratory tests (which may include both safety and efficacy tests), the authors state that ``standard multiplicity adjustment here seems unreasonable, as power will be very low'', and again recommend the use of FDR controlling methods.

All this demonstrates that outside of the context of testing the primary and main secondary endpoints for regulatory approval and labelling, strong FWER control may not be needed, even in confirmatory trials. Less stringent error rates can then be used, where more than one false rejections are acceptable in order to increase the power of the trial. One approach is to control the generalised FWER, or $k$-FWER. The $k$-FWER is the probability of making at least~$k$ false rejections, where $k \geq 1$. Clearly the FWER is a special case of the $k$-FWER when~$k=1$. 
A number of methods controlling the $k$-FWER have been proposed, including step-up procedures~\citep{Lehmann2005, Romano2006, Sarkar2007, Wang2012} and permutation-based procedures~\citep{Romano2007, Romano2008, Romano2010}.
% k-FWER adaptive procedures \citep{Wang2015, Wang2017}
%
% FDP and FDR
Another approach is to accept a certain proportion of false rejections, i.e.\ to control the false discovery proportion (FDP). The FDP is closely related to the well-known false discovery rate (FDR)~\citep{Benjamini1995}, which is now a common error rate to control in experiments with a  large number of hypotheses, such as genomic studies. The FDR is the expected value of the FDP, i.e.\ the FDR is the expected proportion of errors among the rejected hypotheses. Although controlling the FDR controls the expectation of the FDP, in practical applications the actual FDP might be far from its expectation~\citep{Owen2005}. In the context of clinical trials with a relatively small number ($<100$) of hypotheses, this motivates control of the tail probability of the FDP and hence guaranteeing control over the probability of having a high proportion of false discoveries.
Some methods for controlling the FDP have previously been proposed~\citep{Lehmann2005, Romano2008, Romano2006a}.

% Main contribution of paper: extending the graphical approach to controlling the generalised FWER and FDP
In general, the various procedures proposed in the literature for generalised error rate control are not suitable for structured hypothesis testing problems encountered in the context of clinical trials, as they do not respect the underlying hierarchical structure of the testing strategy. In order to do so, in this paper we show how to control both the $k$-FWER and FDP (as well as asymptotic control of the FDR) when using the graphical approach of~\citet{Bretz2009} and its extensions. We achieve this by modifying and applying the methodology for $k$-FWER and FDP control given by~\citet{Laan2004} and~\citet{Romano2010} to the graphical framework. The performance of the resulting procedures are compared analytically and through simulations in the context of various case studies.

% Structure of paper
The rest of the paper is structured as follows. In Section~\ref{sec:framework}, we introduce the basic notation and the graphical approach to hypothesis testing. Section~\ref{sec:kFWER} shows how to modify the graphical approach to control the $k$-FWER, while Section~\ref{sec:graph_FDP} gives a further modification of the graphical approach to control the FDP as well as (asymptotic) control of the FDR. Section~\ref{sec:graph_extensions} shows how to use the proposed procedures for a number of extensions to the graphical approach. We illustrate the proposed methods using three case studies in Section~\ref{sec:case_study}, and conclude with a discussion in Section~\ref{sec:discuss}.

%%%%%%%%%%%%%%%%%%%%%%%%%%%%%%%%%%%%%%%%%%%%%%%%%%%

\section{Graphical approach to hypothesis testing}
\label{sec:framework}

Consider simultaneously testing multiple null hypotheses $H_1, \ldots, H_m$ which are related in some way and so can be thought of as a family of hypothesis tests. Since we are jointly testing multiple hypotheses, there is a resulting multiplicity problem that we wish to take account of in the testing procedure. The standard approach for confirmatory clinical trials is to control the FWER (in the strong sense) below some pre-specified level~$\alpha$, where $\alpha \in (0,1)$. That is, $P(V>0) \leq \alpha$ under any configuration of true and false null hypotheses, where~$V$ denotes the number of false rejections made.
We consider testing $H_1, \ldots, H_m$ using the corresponding $p$-values $p_1, \ldots, p_m$. Let $M = \{1, \ldots, m\}$ denote the associated index set and assume that the $p$-values associated with the true null hypotheses satisfy $P(p_i \leq u) \leq u$  for any $u \in [0,1]$.

We now describe the graphical approach to hypothesis testing introduced by~\citet{Bretz2009}, which controls the FWER. In this approach, the hypotheses $H_1, \ldots, H_m$ are represented by vertices, with associated weights denoting the significance levels. Any two vertices $H_i$ and $H_j$ are connected by a directed edge with weight~$g_{ij}$, which indicates the fraction of the significance level $\alpha_i$ which is propagated from $H_i$ to $H_j$ if $H_i$ is rejected. If $g_{ij} = 0$ then there is no propagation of the significance levels, and the edge can be dropped for convenience from the graphical visualisation. These $g_{ij}$ form an $m \times m$ transition matrix $\bm{G} = (g_{ij})$, which fully characterises the propagation of significance levels.

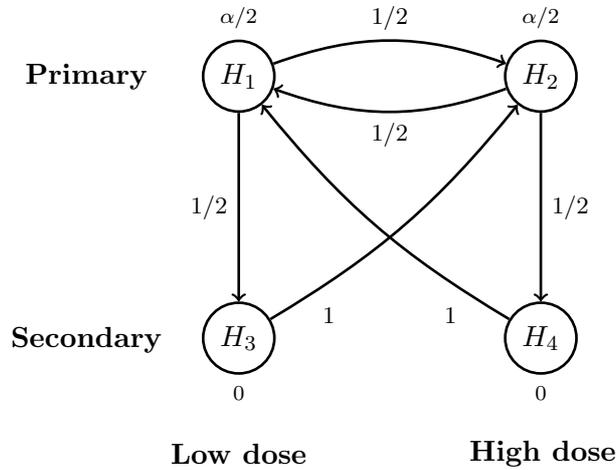
\begin{figure}[ht!]
\centering

%\vspace{12pt}
\begin{tikzpicture}[line width = 1]
% nodes %
\node[circle, draw, text centered, label=above:{\scriptsize $\alpha/2$}] (H1) {$H_1$};
\node[circle, draw, right = 3 of H1, text centered, label=above:{\scriptsize $\alpha/2$}] (H2) {$H_2$};
\node[circle, draw, below=2.5 of H1, text centered, label=below:{\scriptsize $0$}] (H3) {$H_3$};
\node[circle, draw, below=2.5 of H2, text centered, label=below:{\scriptsize $0$}] (H4) {$H_4$};
 
% additional text %
 \node[] at (-2,0) {\textbf{Primary}};
 \node[] at (-2,-3.5) {\textbf{Secondary}};
 \node[] at (0,-5) {\textbf{Low dose}};
 \node[] at (4,-5) {\textbf{High dose}};
   
% edges %
\path[->] (H1) edge [bend left=20] node[above,font=\footnotesize]{$1/2$}  (H2);
\path[->] (H2) edge [bend left=20] node[below,font=\footnotesize]{$1/2$}  (H1);
\draw [->] (H1) -- node[left,font=\footnotesize]{$1/2$}  (H3);
\draw [->] (H2) -- node[right,font=\footnotesize]{$1/2$}  (H4);
\path [->] (H3) edge [bend right=10] node[yshift = -1.2cm, xshift = -1cm, font=\footnotesize]{$1$}  (H2);
\path [->] (H4) edge [bend left=10] node[yshift = -1.2cm, xshift = 1cm,font=\footnotesize]{$1$}  (H1);

\end{tikzpicture} \vspace{12pt}

\caption{Graph showing a possible testing strategy for a diabetes trial with a primary and secondary endpoint that tests two doses of a drug against a placebo, as given in~\citet{Maurer2011}.}

\label{fig:diabetes_trial}

\end{figure}

As an example, consider a trial in diabetes patients that compares two doses (a low dose and a high dose) of an experimental drug against placebo, in terms of both a primary and secondary clinical endpoint. Since the primary endpoint is more important than the secondary one, the trialist tests the primary hypothesis first; only if this is rejected is the secondary hypothesis then tested. Assuming that both doses are equally important, a possible testing strategy is shown in the graph in Figure~\ref{fig:diabetes_trial}, as given in~\citet{Maurer2011}.

%\subsection{Graphical weighting strategies}

\citet{Bretz2011} proposed a graphical weighting strategy which allows the computation of the set of weights for any intersection hypotheses $H_J = \bigcap_{j \in J} \! H_j$, $J \subseteq M$.
%  $m2^{m-1}$ weights for all $2^m - 1$ intersection hypotheses
%A weighted multiple testing procedure can then be applied to each intersection hypothesis $H_J$ to correct for multiplicity, such as a weighted Bonferroni test.
%
The graphical weighting strategy requires the specification of initial weights $w_i(M), i \in M$, for the global null hypothesis $H_{\scriptscriptstyle M}$ and the transition matrix $\bm{G}$, with entries $g_{ij}$ satisfying the regularity conditions
 \begin{equation}
 0 \leq g_{ij} \leq 1, \qquad g_{ii} = 0 \qquad \text{and } \; \sum_{j=1}^m g_{ij} \leq 1 \quad \text{for all } i, j \in M
 \end{equation}
Algorithm~\ref{alg:graph_weight} in Appendix~\ref{Asec:graphFWER} reproduces the algorithm given in~\citet{Bretz2011} for calculating the weights $w_j(J)$, $j \in J$, which can then be used for testing the intersection hypothesis~$H_J$.

Given these weights, a weighted multiple testing procedure can then be applied to each intersection hypothesis~$H_J$, such as a weighted Bonferroni test, or a weighted parametric test if the joint distribution of the $p$-values is known~\citep{Bretz2011}. Applying a weighted Bonferroni test is the simplest option, and leads to the original Bonferroni-based graphical approach for FWER control based on a shortcut procedure where the~$m$ hypotheses can be tested sequentially, and hence requires at most~$m$ steps of the algorithm \citep{Bretz2009} (see also Algorithm~\ref{alg:FWER}).

Adjusted $p$-values can also be calculated when using this graphical approach, which then allow the hypothesis tests to be easily performed at any significance level~$\alpha$. More formally, the adjusted $p$-value $p^{\text{adj}}_j$ for hypothesis~$H_j$ is the smallest significance level at which one can reject the hypothesis using the given multiple test procedure~\citep{Bretz2009}. Algorithm~\ref{alg:adjp} in Appendix~\ref{Asec:graphFWER} reproduces the algorithm given in~\citet{Bretz2009} for calculating adjusted $p$-values.

The R~package \texttt{gMCP}~\citep{Rohmeyer2020} provides functions and a graphical user interface to perform all of the calculations described above.  \vspace{-4pt}

%%%%%%%%%%%%%%%%%%%%%%%%%%%%%%%%%%%%%%%%%%%%%%%%%%%

\section{Graphical approaches for $\lowercase{k}$-FWER control}
\label{sec:kFWER}

%\subsection{Weighted Bonferroni tests for the $k$-FWER}
%\label{subsec:weight_bonf}

Controlling the $k$-FWER at pre-specified level~$\alpha$ implies that $P(V > k) \leq \alpha$, where $V$ is the number of false rejections.
The generalised Bonferroni procedure controls the $k$-FWER \citep{Lehmann2005, Hommel1988}: %\vspace {-6pt}
$\text{Reject any } H_i \text{ for which } p_i \leq k\alpha/m$.
%
% \noindent In multiple testing procedures as in the graphical approach described above, it can be desirable to weight the hypotheses in order to reflect their relative \textit{a-priori} importance. 
%
Assuming known positive weights $w_i$ that satisfy $\sum_{i=1}^m w_i = 1$, \citet{Romano2010} introduced the weighted generalised Bonferroni procedure: $\text{Reject any } H_i \text{ for which } p_i \leq w_i k\alpha.$
%
%This generalised weighted Bonferroni procedure forms the basis for most of the graphical approaches for generalised error rate control in this paper.
If $w_i = 1/m$ for all~$i$ then this is equivalent to the unweighted version. 

In order to extend the graphical approach for controlling the $k$-FWER, it is tempting to simply replace $\alpha$ by $k\alpha$,  in analogy to the modification made for the generalised Bonferroni procedures. However, in general this does not control the $k$-FWER for $k > 1$. As a counterexample, consider the Holm procedure with~$m$ hypotheses, which can be represented as a graph with initial weights $w_i(M) = 1/m$ and $g_{ij} = 1/(m-1)$ for all $i,j \in M$, $i \neq j$. Using the graphical weighting strategy (Algorithm~\ref{alg:graph_weight}), we have $w_i(I) = 1/|I|$ for all $i \in I$ and $I \subseteq M$. Replacing $\alpha$ by $k\alpha$ in the graphical approach (Algorithm~\ref{alg:FWER}) is hence equivalent to a stepdown procedure where the $i$-th smallest $p$-value is compared with the significance level $\alpha_i = \frac{k\alpha}{m+1-i}$. However, since $\frac{k\alpha}{m+1-i} > \frac{k\alpha}{m+k-i}$ for $k > 1$, the result of Theorem 2.3 in~\citet{Lehmann2005} shows that this procedure does not control the $k$-FWER. Hence we turn to alternative procedures for $k$-FWER control.

\subsection{Augmented graphical approach for $k$-FWER control}
\label{subsec:aug_kFWER}

We first consider a simple method of controlling the $k$-FWER described in~\citet[Procedure 1]{Laan2004}, which can be applied to give a graphical approach for $k$-FWER control. The original method starts with an initial procedure that controls the usual FWER, and then augments this by additionally rejecting the hypotheses associated with the smallest $k-1$ remaining (unrejected) $p$-values. These $k-1$ additionally rejected hypotheses can be freely chosen, and so we aim to respect the hierarchical structure of the underlying multiple testing problem and to avoid rejecting hypotheses with large $p$-values. This results in the following augmented graphical approach for $k$-FWER control.

\begin{algorithm}[ -- Augmented graphical approach for $k$-FWER control]\label{alg:kFWER_aug} %\textbf{ -- Augmented graphical approach for $k$-FWER control}
\hspace{0pt} \\[-24pt]
\begin{enumerate}[label=(\roman*)]
\singlespacing
\item Apply the usual Bonferroni-based graphical procedure for FWER control given in Algorithm~\ref{alg:FWER}.
\item Let $I$ denote the index set of any remaining (unrejected) hypotheses. If $I$ is empty then stop; otherwise continue with steps~(ii)--(iv) of Algorithm~\ref{alg:FWER} with $\alpha$ replaced by $\delta$, until up to~$k-1$ additional (augmented) rejections are made. %\vspace{-12pt}
\end{enumerate}

\end{algorithm}

\noindent Here $\delta \geq 0$ determines how many of the `free' rejections we use, and can be set larger than $\alpha$. In fact, we can even set $\delta$ large enough to ensure that $k-1$ additional hypotheses are rejected, regardless of the observed $p$-values (see below). Of course, this comes at the potential cost of rejecting hypotheses with $p$-values close to 1 that are  likely to be null. Conversely, low values of $\delta$ mean that we are only willing to reject a hypothesis if it has reasonably substantial evidence against it. %If $\delta = 0$ then the algorithm reduces to the usual graphical procedure controlling the FWER, since no additional rejections can be made. 

In step~(ii) of Algorithm~\ref{alg:FWER}, there may be a choice as to which of the hypotheses $j \in I$ to reject. Since there can only be a maximum of~$k$ additional rejections in step~(ii) of Algorithm~\ref{alg:kFWER_aug}, the order in which hypotheses are rejected does matter here. One sensible choice is to set $j = \arg \min_{i \in I}\{p_i/w_i(I)\}$, which we use in the remainder of the paper.

The choice of $\delta$ can be data-dependent to ensure that (up to) $k-1$ additional rejections are made. More explicitly, we can increase $\delta$ so that one additional rejection is made, then if necessary increase $\delta$ until another additional rejection is made, and so on. This allows an alternative formulation of the augmented graphical approach based on adjusted $p$-values, which does not depend on an explicit choice of~$\delta$.

\begin{algorithm}[ -- Adjusted augmented graphical approach for $k$-FWER control]\label{alg:kFWER_aug2}
\hspace{0pt} \\[-24pt]
\begin{enumerate}[label=(\roman*)]
\singlespacing
\item Calculate the $m$ adjusted $p$-values $p^{\text{adj}}_i$ corresponding to the usual Bonferroni-based graphical procedure for FWER control, as detailed in Algorithm~\ref{alg:adjp}.
\item Reject all hypotheses $H_i$ with $p^{\text{adj}}_i \leq \alpha$.
\item  Let $I$ denote the index set of any remaining (unrejected) hypotheses. If $I$ is empty then stop; otherwise order the remaining hypotheses in non-decreasing order: $p^{\text{adj}}_{(1)} \leq \cdots \leq p^{\text{adj}}_{(|I|)}$
\item Additionally reject up to $A = \min(|I|, k-1)$ hypotheses $H_{(i)}$ corresponding to $p^{\text{adj}}_{(1)}, \ldots,  p^{\text{adj}}_{(A)}$
\end{enumerate}
\end{algorithm}

\noindent If there are ties in the ordering in step~(iii), they can be broken by choosing the hypothesis with the smallest index, for example. Algorithm~\ref{alg:kFWER_aug} will give the same rejections as Algorithm~\ref{alg:kFWER_aug2} for~$\delta$ large enough. In addition, the R~package \texttt{gMCP}~\citep{Rohmeyer2020} can straightforwardly be used to implement Algorithm~\ref{alg:kFWER_aug} in two stages corresponding to steps~(i) and~(ii).  Hence, we focus on Algorithm~\ref{alg:kFWER_aug} in the rest of the paper. %\\ \vspace{-12pt}

%Unlike the generalised graphical approach, the augmented graphical approach is able to propagate significance levels to all hypotheses that have fewer than~$k$ donors, and in particular is able to reject such hypotheses that have an initial weight of zero. An example of this can be found in Appendix~\ref{Asec:diabetes_trial} for the graph of the diabetes trial given in Figure~\ref{fig:diabetes_trial}. \\ \vspace{-12pt}

\begin{example}[ -- Example of the augmented graphical approach for $k$-FWER control]

\noindent Consider the graph of the diabetes trial given in Figure~\ref{fig:aug_example}, where we control the $k$-FWER for $k=2$ with~$\alpha = 0.05$. Suppose also that the $p$-values are given by $p_1 = 0.01$, $p_2 = 0.03$, $p_3 = 0.02$, $p_4 = 0.024$. \\

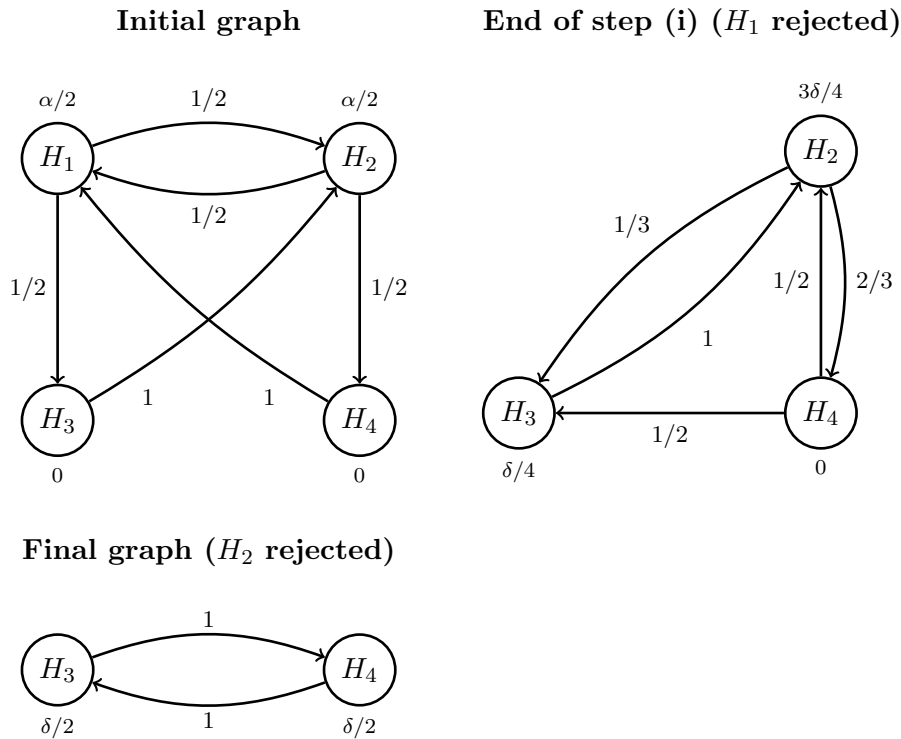
\begin{figure}[ht!]
\centering
\begin{tabular}{c p{12pt} c}

\textbf{Initial graph} & & \textbf{End of step (i) ($H_1$ rejected)} \\[12pt]

\begin{tikzpicture}[line width = 1]
% nodes %
\node[circle, draw, text centered, label=above:{\scriptsize $\alpha/2$}] (H1) {$H_1$};
\node[circle, draw, right = 3 of H1, text centered, label=above:{\scriptsize $\alpha/2$}] (H2) {$H_2$};
\node[circle, draw, below=2.5 of H1, text centered, label=below:{\scriptsize $0$}] (H3) {$H_3$};
\node[circle, draw, below=2.5 of H2, text centered, label=below:{\scriptsize $0$}] (H4) {$H_4$};
 
% additional text %
% \node[] at (2,2) {\textbf{Initial graph}};
   
% edges %
\path[->] (H1) edge [bend left=20] node[above,font=\footnotesize]{$1/2$}  (H2);
\path[->] (H2) edge [bend left=20] node[below,font=\footnotesize]{$1/2$}  (H1);
\draw [->] (H1) -- node[left,font=\footnotesize]{$1/2$}  (H3);
\draw [->] (H2) -- node[right,font=\footnotesize]{$1/2$}  (H4);
\path [->] (H3) edge [bend right=10] node[yshift = -1.2cm, xshift = -1cm, font=\footnotesize]{$1$}  (H2);
\path [->] (H4) edge [bend left=10] node[yshift = -1.2cm, xshift = 1cm,font=\footnotesize]{$1$}  (H1);

\end{tikzpicture}
& &
\begin{tikzpicture}[line width = 1]
% nodes %
\node[circle, draw, right = 3 of H1, text centered, label=above:{\scriptsize $3\delta/4$}] (H2) {$H_2$};
\node[circle, draw, below=2.5 of H1, text centered, label=below:{\scriptsize $\delta/4$}] (H3) {$H_3$};
\node[circle, draw, below=2.5 of H2, text centered, label=below:{\scriptsize $0$}] (H4) {$H_4$};
 
% additional text %
% \node[] at (2,2) {\textbf{After rejecting $H_1$}};
   
% edges %
\path[->] (H2) edge [bend left=15] node[right,font=\footnotesize]{$2/3$}  (H4);
\path[->] (H4) edge [bend left=0] node[left,font=\footnotesize]{$1/2$}  (H2);
\draw [->] (H4) -- node[below,font=\footnotesize]{$1/2$}  (H3);
\path [->] (H3) edge [bend right=15] node[yshift = -0.4cm, xshift = 0.2cm, font=\footnotesize]{$1$}  (H2);
\path [->] (H2) edge [bend right=15] node[yshift = 0.4cm, xshift = -0.2cm, font=\footnotesize]{$1/3$}  (H3);

\end{tikzpicture}

\\[12pt]

\textbf{Final graph ($H_2$ rejected)} &&  \\[12pt]

\begin{tikzpicture}[line width = 1]
 
% nodes %
\node[circle, draw, text centered, label=below:{\scriptsize $\delta/2$}] (H3) {$H_3$};
\node[circle, draw, right=3 of H3, text centered, label=below:{\scriptsize $\delta/2$}] (H4) {$H_4$};

% edges %
\path [->] (H3) edge [bend left=20] node[yshift = 0.2cm, font=\footnotesize]{$1$}  (H4);
\path [->] (H4) edge [bend left=20] node[yshift = -0.2cm,font=\footnotesize]{$1$}  (H3);

\end{tikzpicture}
& & 
%
%\begin{tikzpicture}[line width = 1]
%
%% nodes %
%\node[circle, draw,text centered, yshift = -2cm,  label=below:{\scriptsize $\delta$}] (H4) {$H_4$};
% 
%% additional text %
%% \node[] at (2,2) {\textbf{Final graph}};
%
%\end{tikzpicture}

\vspace{12pt}

\end{tabular}

\caption{The augmented graphical approach for $k$-FWER control applied to the diabetes trial.}

\label{fig:aug_example}

\end{figure}

In step~(i) of Algorithm~\ref{alg:kFWER_aug}, the usual Bonferroni-based graphical procedure for FWER control would only reject $H_1$. The updated graph (i.e.\ removing node $H_1$ and propagating the local significance levels) is then used in step~(ii) of Algorithm~\ref{alg:kFWER_aug}, with $\alpha$ replaced by $\delta$. Supposing that $\delta = 0.5$, we would then reject $H_2$. At this point, we have made $k-1$ additional rejections, and so we stop testing having rejected $H_1$ and $ H_2$. Figure~\ref{fig:aug_example} demonstrates each step of the augmented procedure graphically.

\end{example}

\subsection{Generalised graphical approach for $k$-FWER control}

As an alternative approach, we focus on Algorithm~\ref{alg:graph_weight}, which gives weights $w_j(J), j \in J$, for any $J \subseteq M$. %which can then be used to control the $k$-FWER when testing the multiple hypotheses $H_j, j \in J$. This is achieved by using the weighted generalised Bonferroni procedure: reject any $H_j, j \in J$ for which $p_j \leq w_j(J)k\alpha$.
As shown in~\citet{Bretz2009}, these weights satisfy the monotonicity condition\begin{equation}
\label{eq:monotonicity}
w_j(J) \leq w_j(J') \quad \text{for all } J' \subseteq J \subseteq M \; \text{and } j \in J'.
\end{equation}
Hence, we can apply the generic stepdown method for $k$-FWER control described in~\citet[Algorithm 4.1]{Romano2010} with these weights to give the following generalised Bonferroni-based algorithm. Essentially, we simply set the critical constants $\hat{c}_{n,K,i}(1- \alpha, k)$ in their algorithm equal to $w_i(K)$, where $K$ is used in step~(iv) of Algorithm~\ref{alg:kFWER}  below to index the subsets including $k - 1$ of the previously rejected hypotheses. In what follows, we refer to this as the generalised graphical approach for $k$-FWER control.

\begin{algorithm}[ -- Generalised graphical approach for $k$-FWER control]\label{alg:kFWER}
\hspace{0pt} \\[-24pt]
\begin{enumerate}[label=(\roman*)]
\singlespacing
\item Set $I = M$.
\item Reject any $H_i$, $i \in I$ for which $p_i \leq w_i(I)k\alpha$
\item Let $R = \{i \in I : p_i \leq w_i(I)k\alpha \}$. If $|R| < k$ or $|R| = |I|$ then stop; otherwise update $I \rightarrow I \setminus R$.
\item Reject any $H_i$, $i \in I$ for which \[
p_i \leq \min_{J \subseteq R, |J| = k-1} \{w_i(K) : K = I \cup J \} k \alpha
\] If no such $H_i$ exists then stop; otherwise let $R'$ be the indices of these rejected hypotheses.
\item Update the sets $I$ and $R$ as follows: \vspace{-6pt}\begin{align*}
I & \rightarrow I \setminus R'\\
R & \rightarrow R \cup R' 
\end{align*}
\item If $|I| \geq 1$, go to step (iv); otherwise stop.
\end{enumerate}

\end{algorithm}

\noindent In Algorithm~\ref{alg:kFWER}, at each step $R$ is simply the set of indices of all the hypotheses that have been rejected previously, and $I$ is the set of indices of the remaining hypotheses $M \setminus R$. The algorithm is in a similar spirit to the graphical weighting strategy \citep{Bretz2011}, in the sense that there is a separation between the weighting strategy and the graphical test procedure which allows the generalisation to $k$-FWER control. 

In step~(iii) of Algorithm~\ref{alg:kFWER}, if $|R| < k-1$ and $|R| \neq |I|$ then we can freely reject additional hypotheses so that a total of (up to) $k-1$ rejections are made, while still controlling the $k$-FWER, since the algorithm will stop at this step. In order to respect the hierarchical structure of the underlying multiple testing procedure, and to avoid rejecting hypotheses with large $p$-values, we propose the following sub-procedure in step~(iii) if $|R| < k-1$: \vspace{-12pt}
\begin{enumerate}
\singlespace
\item Set $I \rightarrow I \setminus R$ and follow steps (ii)--(iv) of the usual Bonferroni-based graphical procedure for FWER control (Algorithm~\ref{alg:FWER}) with~$\alpha$ replaced by~$\delta$, until up to $k-1$ additional rejections have been made.
\end{enumerate}
\noindent As before, $\delta \geq 0$ determines how many of the `free' rejections we use, and hence can be set larger than $\alpha$ or made data-dependent so that (up to) $k-1$ additional rejections are made.

Looking at Algorithm~\ref{alg:kFWER} as a whole, if $k =1$, then once a hypothesis is rejected, it no longer plays a further role and step~(iv) above reduces to rejecting any $H_i$, $i \in I$, for which $p_i \leq w_i(I)\alpha$. Hence, Algorithm~\ref{alg:kFWER} is equivalent to Algorithm~\ref{alg:FWER} (the usual Bonferroni-based graphical approach for FWER control) in that both algorithms will lead to exactly the same rejections when $k = 1$, assuming the same initial weights.
When $k > 1$, however, the algorithm becomes more complex and involves maximising over subsets including $k-1$ of the previously rejected hypotheses in step~(iv). As noted by~\citet{Romano2010}, intuitively this is because when considering a set of unrejected hypothesis in Algorithm~\ref{alg:kFWER}, we may have already rejected (hopefully at most) $k-1$ true null hypotheses. We do not know which of the rejected hypotheses are true, and so we maximise over subsets including at most $k-1$ of those hypotheses previously rejected. % As shown in the rest of the paper, this has additional negative consequences when the testing procedure has a hierarchical structure.
In Appendix~\ref{Asubsec:streamlined} we discuss the computational challenges of using Algorithm \ref{alg:kFWER} for large values of $m$, and show how to streamline and operationalise the algorithm. However, in general these modified procedures only give asymptotic control of the $k$-FWER as the sample size of the trial increases.

In Appendix \ref{Asubsec:example_kgMCP} we give some examples of using the generalised graphical approach. We show how it reduces to previous algorithms for $k$-FWER control as special cases, but also how it can have undesirable properties when the testing procedure has a hierarchical structure. The main problem (as demonstrated analytically in Example~4 of Appendix~\ref{Asubsec:example_kgMCP}) is that if a hypothesis~$H_j$ has fewer than~$k$ donors, its initial significance level will never increase, except for up to $k-2$ hypotheses via the sub-procedure in step~(iii).  Here, the donors of a hypothesis~$H_j$ are the hypotheses that donate (or propagate) their significance levels to $H_j$ if they are rejected.
%More formally, we denote the donors of hypothesis~$H_j$ by $\text{do}(H_j) = \{H_i : g_{ij}>0\}$.
Hence, the generalised graphical approach cannot effectively propagate the significance levels through the graph.  We will see further examples of this in the case studies given in Section~\ref{sec:case_study}.

\begin{example}[ -- Example of the generalised graphical approach for $k$-FWER control]

\noindent We again consider the graph of the diabetes trial given in Figure~\ref{fig:diabetes_trial}. We control the $k$-FWER for $k=2$ with~$\alpha = 0.05$, with the $p$-values this time given by $p_1 = 0.01$, $p_2 = 0.03$, $p_3 = 0.02$, $p_4 = 0.024$. Applying the generalised graphical approach for $k$-FWER control gives the following:

\begin{enumerate}%[label=(\roman*)]
\singlespacing
\item Set $I = \{1,2,3,4\}$.
\item We reject any $H_i$, $i \in I$, for which $p_i \leq w_i(I)k\alpha$. Here $w_i(I)$ are simply the initial weights and so $w_1(I) = w_2(I) = 0.5$ and $w_3(I) = w_4(I) = 0$. Since $p_1< \alpha$, $p_2 < \alpha$,  $H_1$ and $H_2$ are rejected at this step.
\item We reject $H_i$, $i \in I = \{3, 4 \}$, if $p_i \leq \min \{ w_i(\{1,3,4\}), w_i(\{2,3,4\}) \}$. However, $w_3(\{1,3,4\}) = w_4(\{2,3,4\}) = 0$ and hence neither $H_3$ nor $H_4$ can be rejected.
\end{enumerate}

\end{example}

%\noindent In contrast, as shown in Section \ref{subsec:aug_kFWER}, if we use the augmented graphical approach for $k$-FWER control then we can additionally reject $H_3$.

\subsection{Existing power comparisons}

\citet{Romano2007} argue that the augmented procedure is suboptimal compared with their generic stepdown method for $k$-FWER control, since it can only reject at most $k-1$ hypotheses more compared to a usual FWER-controlling procedure, whereas Algorithm~\ref{alg:kFWER} can reject substantially more hypotheses. In their simulation study~\citep[Section 6]{Romano2007}, they considered testing the means of a multivariate normal distribution with common correlation~$\rho$, where the number of hypotheses $M = 50$ or $M = 400$. They compared a number of different procedures for $k$-FWER control, but the relevant power comparison for our context of graphical approaches is the one between the generalised Holm procedure and the augmented Holm procedure. Their simulation results showed that when $M = 400$, $k = 10$ and $\rho \leq 0.5$, the generalised Holm procedure can make a substantially higher number of rejections (up to twice as many) compared with the augmented Holm procedure. However, when $M = 50$ and $k = 3$, the augmented Holm procedure almost always had a higher number of rejections than the generalised Holm procedure.

These findings are corroborated by the simulation results of~\citet{Dudoit2004}. They also considered testing the means of a multivariate normal distribution, with the number of hypotheses $M = 24$ or $M = 400$.  Through simulation, they compared the augmented and generalised Holm and Bonferroni procedures, concluding that the augmented approach tends to be more powerful than the generalised approach ``for a broad range of models''\citep[Section 6.2.1]{Dudoit2004}. The largest gains in power were when the number of hypotheses was small and a large proportion of the null hypotheses were true. However, for a large number of hypotheses ($M = 400$) and when $\alpha$ was relatively large, the generalised approaches was more powerful than the augmented approaches. In many clinical trials, we would be in the setting with a smaller number of hypotheses, and so the augmented approach would be expected to be more powerful. In our case studies in Section~\ref{sec:case_study}, we consider power comparisons beyond Bonferrroni or Holm based methods. %we return to this issue in the Discussion (Section~\ref{sec:discuss}).

%%%%%%%%%%%%%%%%%%%%%%%%%%%%%%%%%%%%%%%%%%%%%%%%%%%%

\section{Graphical approaches for FDP and (asymptotic) FDR control}
\label{sec:graph_FDP}

In this section, we consider how to extend the graphical approach for FDP and (asymptotic) FDR control. More formally, the FDP is defined as $\text{FDP} = \frac{V}{\max(R,1)}$, where~$R$ denotes the total number of rejections. The FDR is then the expectation of the FDP. A multiple testing procedure controls the tail probability of the FDP at level~$\alpha$ if $P(\text{FDP} > \gamma) \leq \alpha$, where $\gamma \in [0,1)$ is a pre-specified bound. This is also known as the tail probability for the proportion of false positives \citep{Dudoit2004} or the false discovery exceedance \citep{Javanmard2018}. Note that setting~$\gamma = 0$ results in control of the FWER at level~$\alpha$. In what follows, when we refer to FDP control, we mean controlling this tail probability of the FDP, where we suppress the dependence on~$\gamma$ for notational convenience. 

\subsection{Augmented approach for FDP and FDR control}

A simple method of controlling the FDP based on a FWER-controlling procedure is given by~\citet[Procedure~2]{Laan2004}. This can be applied to give an augmented graphical approach for FDP control, in a similar way to that for $k$-FWER control. A proof that FDP control holds can be found in~\citet[Theorem~2]{Laan2004}.

%\newpage

\begin{algorithm}[ -- Augmented graphical approach for FDP control]\label{alg:FDP_aug}
\hspace{0pt} \\[-24pt]
\begin{enumerate}[label=(\roman*)]
\singlespacing
\item Apply the usual Bonferroni-based graphical FWER procedure given in Algorithm~\ref{alg:FWER}. Let~$R$ denote the index set of the rejected hypotheses.
\item Let $I$ denote the index set of any remaining (unrejected) hypotheses. If $I$ is empty, then stop.

\item Let $D$ be the largest integer satisfying

\[
\frac{D}{D+|R|} \leq \gamma
\] \vspace{0pt}

If $D = 0$ then stop; otherwise continue with steps (ii)--(iv) of Algorithm~\ref{alg:FWER} with $\alpha$ replaced by $\delta$, until up to $D$ additional (augmented) rejections are made.

\end{enumerate}
\end{algorithm}

\noindent Here $\delta \geq 0$ is a constant controlling how many additional rejections are made. As before, $\delta$ may be greater than $\alpha$, and can be set very large so that all~$D$ additional hypotheses are rejected.
The choice of $\delta$ can also be data-dependent, giving an alternative algorithm based on adjusted $p$-values, which does not depend on an explicit choice of~$\delta$.

\begin{algorithm}[ -- Adjusted augmented graphical approach for FDP control] \label{alg:FDP_aug2}
\hspace{0pt} \\[-24pt]
\begin{enumerate}[label=(\roman*)]
\singlespacing
\item Calculate the $m$ adjusted $p$-values $p^{\text{adj}}_i$ corresponding to the usual Bonferroni-based graphical procedure for FWER control, as detailed in Algorithm~\ref{alg:adjp}.
\item Reject all hypotheses $H_i$ with $p^{\text{adj}}_i \leq \alpha$.
\item  Let $I$ denote the index set of any remaining (unrejected) hypotheses. If $I$ is empty then stop; otherwise order the remaining hypotheses in non-decreasing order: $p^{\text{adj}}_{(1)} \leq \cdots \leq p^{\text{adj}}_{(|I|)}$
\item Additionally reject up to $A = \min(|I|, D)$ hypotheses $H_{(i)}$ corresponding to $p^{\text{adj}}_{(1)}, \ldots,  p^{\text{adj}}_{(A)}$

\end{enumerate}
\end{algorithm}

\noindent If there are ties in the ordering in step~(iii), they can be broken by choosing the hypothesis with the smallest index. Algorithm~\ref{alg:FDP_aug} will give the same rejections as Algorithm~\ref{alg:FDP_aug2} for $\delta$ large enough. Also, the R~package \texttt{gMCP}~\citep{Rohmeyer2020} can straightforwardly be used to implement Algorithm~\ref{alg:FDP_aug} in two stages corresponding to steps~(i) and~(iii). Hence, we focus on Algorithm~\ref{alg:FDP_aug} in the remainder of the paper.

\begin{example}[ -- Example of the augmented graphical approach for FDP control]

We continue the example of the diabetes trial displayed in Figure~\ref{fig:diabetes_trial}, where this time we aim to control the FDP with~$\alpha = 0.05$ and $\delta = 0.5$. Suppose this time the $p$-values are given by $p_1 = 0.01$, $p_2 = 0.015$, $p_3 = 0.02$, $p_4 = 0.024$. 
In step (i), the Bonferroni-based graphical procedure for FWER control would reject $H_1$ and $H_2$. We then reject up to $D$ additional hypotheses in step (iii), where $D$ is the largest integer satisfying $D/(D+2) \leq \gamma$. Hence if $0 \leq \gamma < 1/3$ we make $D = 0$ additional rejections, if $1/3 \leq \gamma < 1/2$ we make $D = 1$ additional rejection (reject $H_3$), and if $\gamma \geq 1/2$ we make $D=2$ additional rejections (reject $H_3$ and $H_4$). 

\end{example}

\noindent Although our focus in this paper is on controlling the tail probability of the FDP, we note in passing that the augmented procedure for FDP control at level $\alpha$ automatically gives asymptotic control of the FDR at level $2 \alpha$. This follows directly from~\citet[Theorem~3]{Laan2004}. Hence, applying the augmented graphical approach for FDP control given in Algorithm~\ref{alg:FDP_aug} at pre-specified level~$\alpha$ asymptotically controls the FDR at level $2\alpha$.
%
%\begin{algorithm} \textbf{-- Augmented graphical approach for asymptotic FDR control}
%\label{alg:FDR_aug}
%
% \end{algorithm}
%
\citet{Lehmann2005} showed that FDP control at level $\alpha$ also implies FDR control at level $\alpha^* = \alpha(1-\gamma)+\gamma$. Hence, if $\alpha^* < 2\alpha$, which implies that  $\gamma < \alpha/(1 - \alpha)$, this bound can be used instead, while also yielding finite sample FDR control.

\subsection{Generalised graphical approach for asymptotic FDP control}

As an alternative method to control the FDP, we can directly apply the generic method for FDP control in~\citet[Algorithm~8.1]{Romano2010} to give the following graphical approach. 

\begin{algorithm}[ -- Generalised graphical approach for asymptotic FDP control] \label{alg:FDP}
\hspace{0pt} \\[-24pt]
\begin{enumerate}[label=(\roman*)]
\singlespacing
\item Let $j=1$ and $k_1 = 1$
\item Apply the $k_j$-FWER procedure given in Algorithm~\ref{alg:kFWER}, and let $R_j$ denote the index set of the hypotheses it rejects.
\item If $|R_j| < k_j/\gamma - 1$, stop and reject all hypotheses rejected by the $k_j$-FWER procedure. Otherwise, let $j = j+1$ and $k_j = k_{j-1} + 1$, then return to step~(ii).

\end{enumerate}
\end{algorithm}

\noindent This algorithm was only proven in~\citet{Romano2010} to give asymptotic FDP control, but they showed empirically that it had good finite control of the FDP. However, since Algorithm~\ref{alg:FDP} is based on the $k$-FWER generalised graphical approach, the same potential problems as described in Appendix~\ref{Asubsec:example_kgMCP} will also apply. Finally, we again note in passing that the result of~\citet{Lehmann2005} shows that this procedure gives (asymptotic) FDR control at level $\alpha^* = \alpha(1-\gamma)+\gamma$. 

\begin{example}[ -- Example of the generalised graphical approach for FDP control]

We continue the example of the diabetes trial displayed in Figure~\ref{fig:diabetes_trial}, where we aim to control the FDP with~$\alpha = 0.05$ and $\delta = 0.5$. Suppose the $p$-values are given by $p_1 = 0.01$, $p_2 = 0.015$, $p_3 = 0.02$, $p_4 = 0.024$. In step~(ii), applying the FWER procedure results in the rejection of $H_1$ and $H_2$. Hence $|R_1| = 2$ and we stop if $\gamma < 1/3$. If $\gamma \geq 1/3$, then we apply the 2-FWER procedure, which again rejects $H_1$ and $H_2$. Hence $|R_2| = 2$ and we stop if $\gamma < 2/3$.  If $\gamma \geq 2/3$, then we apply the 3-FWER procedure, which rejects $H_1$ and $H_2$. Since $|R_3| = 2 < 3/\gamma - 1$ for all $\gamma < 1$ we would stop at this step.

\end{example}

\subsection{Existing power comparisons}

\citet{Romano2007} argue that the augmented procedure for FDP control is suboptimal compared with their generalised method for FDP control, given that both are based on the $k$-FWER controlling procedures. In the simulation results for FDP controlling procedures given in~\citet{Dudoit2004} and~\citet{Romano2007}, the augmented and generalised approaches as given above (Algorithms~\ref{alg:FDP_aug} and~\ref{alg:FDP}) are not directly compared for Holm (or Bonferroni) based procedures. However, given their simulation results for $k$-FWER control, we might also expect the augmented approach to have a higher power than the generalised approach when the number of hypotheses are small or when the proportion of true null hypotheses is high. We consider such power comparisons in our case studies in Section~\ref{sec:case_study}. %, and return to this issue in the Discussion (Section~\ref{sec:discuss}).
%\newpage

%%%%%%%%%%%%%%%%%%%%%%%%%%%%%%%%%%%%%%%%%%%%%%%%%%%%
%\newpage

\section{Extensions to the graphical approach}
\label{sec:graph_extensions}

\noindent The original Bonferroni-based graphical approach of~\citet{Bretz2009} has been extended in a number of ways~\citep{Bretz2014}.  These extensions can be used in the augmented and generalised procedures for $k$-FWER and FDP control. 

\subsection{Entangled graphs}
\label{subsec:entangled}

\noindent Firstly we consider the setting where it is desirable for the graphical procedures to have memory, in the sense that the propagation of significance levels depends on their origin. To achieve this, we can define individual graphs for each relationship and combine them afterwards. This is known as an entangled graph, and the algorithm presented in~\citet{Maurer2013} gives an entangled Bonferroni-based graphical approach.

%was described by~\citet{Maurer2013} as follows.
%Consider~$n$ graphs that we wish to entangle, denoted by $\mathcal{G}_h (\bm{w}_h, \bm{G}_h)$ for $h = 1, \ldots, n$. Here $\bm{G}_h$ are the $m \times m$ transition matrices. 

Hence, we can straightforwardly modify the augmented graphical approaches for $k$-FWER and FDR control for use with entangled graphs. To do so, simply replace Algorithm~\ref{alg:FWER} with the algorithm of~\citet{Maurer2013}. For the adjusted augmented graphical approaches, replace Algorithm~\ref{alg:adjp} with the algorithm of~\citet{Maurer2014}, which shows how to calculate adjusted $p$-values for the entangled graph setting.
\citet{Maurer2013} also showed how to calculate the weights for any intersection hypothesis~$H_J$, $J \subseteq M$, and this weighting strategy satisfies the monotonicity condition given in equation~\eqref{eq:monotonicity}. Hence, we can directly apply this weighting strategy to the generalised graphical approaches for $k$-FWER and FDP control. We give an example of the use of entangled graphs in the case study described in Section~\ref{subsec:Pre-RELAX}.

\subsection{Weighted parametric tests}

\noindent All the procedures so far have been based on weighted Bonferroni tests, which can be conservative. As an alternative, weighted parametric tests can be used if the joint distribution of the $p$-values $p_j$, $j \in J$, are known for the intersection hypothesis~$H_J$. In this case, a weighted min-$p$ test can be defined~\citep{Westfall1993, Westfall1998}. This test rejects~$H_J$ if there exists a $j \in J$ such that $p_j \leq c_J w_j(J) \alpha$, where $c_J$ is the largest constant satisfying
\[
P_{H_J} \left( \bigcup_{j \in J} \{ p_j \leq c_j w_j(J) \alpha\} \right) \leq \alpha.
\]
\noindent If only some of the multivariate distributions of the $p$-values are known, then~\citet{Bretz2011} and~\citet{Xi2017} showed how to derive conservative upper bounds on this rejection probability, and hence determine a value for $c_I$.

The motonocity condition in this setting is \begin{equation}
\label{eq:monotonicity_param}
c_J w_j(J) \leq c_{J'} w_j(J') \quad \text{for all } J' \subseteq J \subseteq M \; \text{and } j \in J'.
\end{equation}

\noindent which implies that rejection thresholds are always more liberal when fewer hypotheses are included in the set.
In practice, this condition is often violated when using weighted parametric tests~\citep{Bretz2011}. If this is the case, then it may be possible to modify the weighting scheme so that equation~\eqref{eq:monotonicity_param} holds~\citep{Bretz2011, Xi2017}.
If the monotonicity condition does hold, then we can use the weighted parametric tests directly for the augmented and generalised approaches for $k$-FWER and FDP control, with the only change being that $w_i(I)$ is replaced by $c_I w_i(I)$. For the adjusted augmented graphical approach, adjusted $p$-values can be constructed for weighted parametric tests~\citep{Xi2017}.

\subsection{Group sequential designs}

\noindent The graphical approach can also be extended to group sequential designs with one or more interim analyses. Under mild monotonicity conditions,\citet{Maurer2013a} proposed a graphical testing procedure for multiple hypotheses and multiple interim analyses. % Hence, like before we can directly apply this weighting strategy to both the augmented and the generalised graphical approaches for $k$-FWER control.  We refer the interested reader to~\citet{Maurer2013a} for further details.
More formally, consider testing $H_1, \ldots, H_m$ in a group sequential trial at time points $t = 1, \ldots, h$.  Each $H_i$ has an associated error spending function $a_i (\kappa ,y)$ with information fraction $y$ and significance level $\kappa$. The nominal significance levels are denoted by $\tilde{\alpha}_{i,t} (\kappa)$, which are the interim decision boundaries. We assume that these nominal levels satisfy the monotonicity condition $\tilde{\alpha}_{i,t}(\kappa') \geq \tilde{\alpha}_{i,t}(\kappa)$ for all $\kappa' > \kappa$ (i.e.\ the rejection boundaries are always higher when the total error rate of the design is higher). These conditions hold for many spending functions, including O'Brien-Fleming and Pocock boundaries~\citep{Maurer2013a}.
% spent levels $\alpha_{i,t} = a_i  (\gamma ,y_t) - a_i (\gamma, y_{t-1})$
The algorithm presented in~\citet{Maurer2013a} gives a Bonferroni-based graphical test procedure for group sequential designs.

The augmented graphical approaches for $k$-FWER and FDP control can hence be extended to apply to group sequential designs: simply replace Algorithm~\ref{alg:FWER} with the algorithm in~\citet{Maurer2013a}. For the adjusted augmented graphical approach, replace Algorithm~\ref{alg:adjp} with the algorithm of~\citet{Maurer2014}, which shows how to calculate adjusted $p$-values for the group sequential design setting.

%%%%%%%%%%%%%%%%%%%%%%%%%%%%%%%%%%%%%%%%%%%%%%%%%%%%

\section{Case Studies}
\label{sec:case_study}

In this section, we compare and contrast the use of the algorithms for $k$-FWER and FDP control on three clinical case studies covering a broad range of clinical trial applications. In Section~\ref{subsec:pharmaco} we revisit an exploratory pharmacodynamic clinical trial to investigate the effect of drug activity at the GABA-A receptor in the brain. In Section~\ref{subsec:Pre-RELAX} we revisit a proof-of-concept trial investigating three doses of a new drug against a placebo on multiple biological endpoints related to acute heart failure. Finally, in Section~\ref{subsec:atmosphere} we illustrate the proposed approaches for the comparison of three therapies in a confirmatory clinical trial for heart failure patients.

\subsection{Pharmacodynamic study}
\label{subsec:pharmaco}

\begin{figure}[ht!]
\centering

\begin{tikzpicture}[line width = 1]

\scriptsize

% nodes %
\node (T1D1) [circle,  draw,  align=center]  {$T_1 D_1$\\0};
\node (T1D2) [circle,  draw,  right=3 of  T1D1, align=center]  {$T_1 D_2$\\0};
\node (T1D3) [circle,  draw,  right=3 of  T1D2, align=center]  {$T_1 D_3$\\0};
\node (T2D1) [circle,  draw,  below=3 of  T1D1, align=center]  {$T_2 D_1$\\0};
\node (T2D2) [circle,  draw,  right=3 of  T2D1, align=center]  {$T_2 D_2$\\0};
\node (T2D3) [circle,  draw,  right=3 of  T2D2, align=center]  {$T_2 D_3$\\0};
\node (T3D1) [circle,  draw,  below=3 of  T2D1, align=center]  {$T_3D_1$\\0};
\node (T3D2) [circle,  draw,  right=3 of  T3D1, align=center]  {$T_3 D_2$\\0};
\node (T3D3) [circle,  draw,  right=3 of  T3D2, align=center]  {$T_3 D_3$\\0};
\node (T4D1) [circle,  draw,  below=3 of  T3D1, align=center]  {$T_4D_1$\\0};
\node (T4D2) [circle,  draw,  right=3 of  T4D1, align=center]  {$T_4 D_2$\\0};
\node (T4D3) [circle,  draw,  right=3 of  T4D2, align=center]  {$T_4 D_3$\\$\alpha/3$};
\node (T5D1) [circle,  draw,  below=3 of  T4D1, align=center]  {$T_5D_1$\\0};
\node (T5D2) [circle,  draw,  right=3 of  T5D1, align=center]  {$T_5 D_2$\\$\alpha/3$};
\node (T5D3) [circle,  draw,  right=3 of  T5D2, align=center]  {$T_5 D_3$\\$\alpha/3$};

% additional text %
{\large
 \node[] at (4.25,1.75) {\textbf{Dose}};
 \node[ rotate=90] at (-2,-8.25) {\textbf{Time}};
 }

% edges %

\draw [->] (T1D2) -- node[above]{$1/2$}  (T1D1);
\draw [->] (T1D3) -- node[above]{$1/2$}  (T1D2);
\draw [->] (T5D2) -- node[above]{$1/2$}  (T5D1);
\draw [->] (T5D3) -- node[above]{$1/2$}  (T5D2);

\draw [->] (T2D1) -- node[left]{$1/2$}  (T1D1);
\draw [->] (T3D1) -- node[left]{$1/2$}  (T2D1);
\draw [->] (T4D1) -- node[left]{$1/2$}  (T3D1);
\draw [->] (T5D1) -- node[left]{$1/2$}  (T4D1);

\draw [->] (T2D3) -- node[right]{$1/2$}  (T1D3);
\draw [->] (T3D3) -- node[right]{$1/2$}  (T2D3);
\draw [->] (T4D3) -- node[right]{$1/2$}  (T3D3);
\draw [->] (T5D3) -- node[right]{$1/2$}  (T4D3);

\path[->] (T2D1) edge [bend left=15] node[xshift=-0.25cm, yshift=0.25cm]{$1/2$}  (T1D2);
\path[->] (T1D2) edge [bend left=15] node[xshift=0.25cm, yshift=-0.25cm]{$1/2$}  (T2D1);
\path[->] (T2D2) edge [bend left=15] node[xshift=-0.25cm, yshift=0.25cm]{$1/2$}  (T1D3);
\path[->] (T1D3) edge [bend left=15] node[xshift=0.25cm, yshift=-0.25cm]{$1/2$}  (T2D2);
\path[->] (T3D1) edge [bend left=15] node[xshift=-0.25cm, yshift=0.25cm]{$1/2$}  (T2D2);
\path[->] (T2D2) edge [bend left=15] node[xshift=0.25cm, yshift=-0.25cm]{$1/2$}  (T3D1);
\path[->] (T3D2) edge [bend left=15] node[xshift=-0.25cm, yshift=0.25cm]{$1/2$}  (T2D3);
\path[->] (T2D3) edge [bend left=15] node[xshift=0.25cm, yshift=-0.25cm]{$1/2$}  (T3D2);
\path[->] (T4D1) edge [bend left=15] node[xshift=-0.25cm, yshift=0.25cm]{$1/2$}  (T3D2);
\path[->] (T3D2) edge [bend left=15] node[xshift=0.25cm, yshift=-0.25cm]{$1/2$}  (T4D1);
\path[->] (T4D2) edge [bend left=15] node[xshift=-0.25cm, yshift=0.25cm]{$1/2$}  (T3D3);
\path[->] (T3D3) edge [bend left=15] node[xshift=0.25cm, yshift=-0.25cm]{$1/2$}  (T4D2);
\path[->] (T5D1) edge [bend left=15] node[xshift=-0.25cm, yshift=0.25cm]{$1/2$}  (T4D2);
\path[->] (T4D2) edge [bend left=15] node[xshift=0.25cm, yshift=-0.25cm]{$1/2$}  (T5D1);
\path[->] (T5D2) edge [bend left=15] node[xshift=-0.25cm, yshift=0.25cm]{$1/2$}  (T4D3);
\path[->] (T4D3) edge [bend left=15] node[xshift=0.25cm, yshift=-0.25cm]{$1/2$}  (T5D2);

\end{tikzpicture} \vspace{12pt}

\caption{Graph representing the testing strategy for the pharmacodynamic study described in~\citet{Ferber2011}, with modified initial weights. Here contrast $T_i D_j$ compares the change from baseline under dose $j$ ($j = 1,2,3$) at time point $i$ ($i = 1, \ldots, 5$) to the corresponding change under placebo.}
\label{fig:PD_graph}
\end{figure}
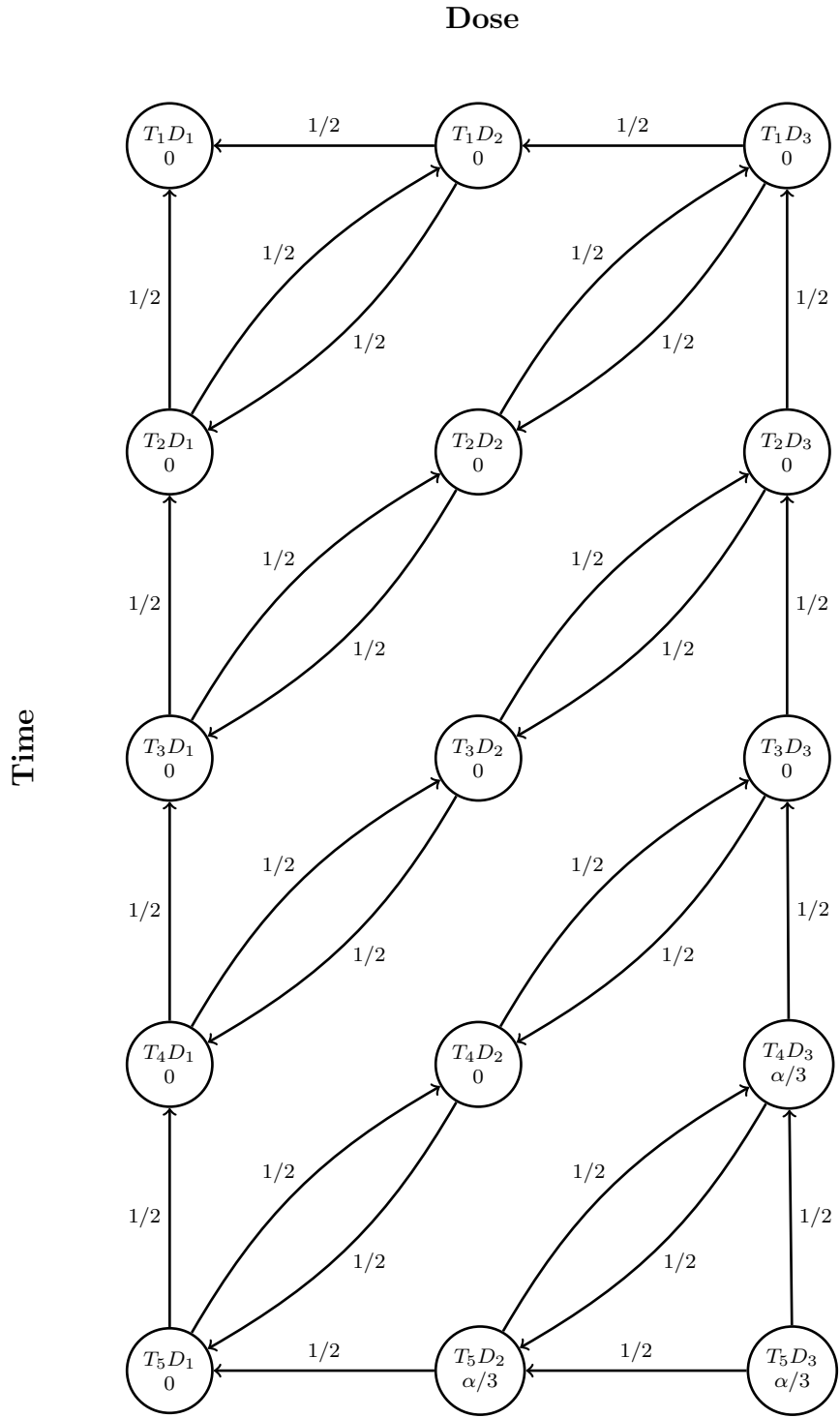

Our first case study is motivated by the exploratory pharmacodynamic clinical study reported by~\citet{Ferber2011}, which explored the effect of drug activity at the GABA-A receptor in the brain as measured using a quantitative electroencephalogram (qEEG). Three doses of the drug (0.25mg, 0.5mg and 1mg) were tested as well as a placebo. During the first 15 minutes after the drug was given to each patient, qEEG measurements were taken and afterwards subdivided into 5 time slices of 3 minutes duration.
The analysis strategy used a mixed effect linear model to obtain~15 contrasts to formally test. Contrast $T_i D_j$ compared the change from baseline under dose $j$ ($j = 1,2,3$) at time point $i$ ($i = 1, \ldots, 5$) to the corresponding change under placebo. Figure \ref{fig:PD_graph} shows the graph representing the hierarchical testing strategy used for these 15 hypotheses (with modified initial weights, see below), and Table \ref{tab:PD_pval} gives the unadjusted $p$-values from the mixed effects linear model for the~15 hypotheses.

\begin{table}[H]
\centering

\begin{tabular}{p{1.5cm} p{0cm} p{1.8cm} p{2cm} p{2cm} p{2cm} p{2.2cm}}
\\ \hline \Tstrut
 & & \multicolumn{5}{c}{\textbf{Time}} \\ \cline{3-7}
\Tstrut  \textbf{Dose} & & $T_1$ & $T_2$ & $T_3$ & $T_4$ & $T_5$ \\ \hline
$D_1$ & & 0.7808 & 0.0600 & 0.0137 & 0.0724 & 0.0162 \\
$D_2$ & & 0.9433 & 0.0053 & $6.5 \times 10^{-6}$ & $2.8 \times 10^{-6}$ & $9.1 \times 10^{-8}$\\
$D_3$ & & 0.9993 & $1.0 \times 10^{-5}$ & $1.7 \times 10^{-11}$ & $4.2 \times 10^{-12}$ & $8.1 \times 10^{-13}$
 \\ [1ex] \hline

\end{tabular}

\caption{\label{tab:PD_pval} \textit{Table of $p$-values for the pharmacodynamic study of~\citet{Ferber2011}.}}

\end{table}

Figure \ref{fig:PD_graph} shows that the hypotheses $T_4 D_3$ and $T_5 D_2$ each only have a single donor hypothesis ($T_5 D_3$). Hence if they have initial weights of zero (as in the original graph~\citep{Ferber2011}), then they cannot be rejected by the generalised graphical approach for $k$-FWER control with $k = 2$. This then means that no hypotheses can be rejected except for $T_5 D_3$. Therefore, we first set the initial weights for $T_4 D_3, T_5 D_2$ and $T_5 D_3$ to 1/3, with all other weights set equal to zero.

Table~\ref{tab:PD_rej1}  shows the resulting rejections for the generalised and augmented graphical $k$-FWER and FDP controlling procedures, with $\delta=1$. 
Looking first at the $k$-FWER procedures, for $k=1$ the generalised and augmented graphical procedures both reject the same~8 hypotheses, as would be expected. For $k=2$ and $k=3$, the augmented procedure rejects~9 and~10 hypotheses respectively. However, the generalised graphical procedure rejects fewer hypotheses when $k =2$ and $k=3$, with only 3 rejections in both cases. For $k=3$ this is because all hypotheses have fewer than~3 donors and hence only those hypotheses with non-zero initial weights can be rejected. This is still the case when $k=2$, even though all hypotheses (except for $T_5 D_3$) have 2 donors, showing that the generalised graphical procedure cannot effectively propagate the significance levels through the graph.

\begin{table}[ht!]
\centering
\doublespacing
\begin{tabular}{l l p{0pt}l }
\\ \hline \Tstrut
\textbf{Procedure} & & & \textbf{Rejected hypotheses} \\ \hline
$k$-FWER \\ \cline{1-2}
\multirow{3}{*}{Generalised} & $k=1$ & & $T_2 D_3, T_3 D_2, T_3 D_3, T_4 D_2, T_4 D_3, T_5 D_1, T_5 D_2, T_5 D_3$ \\
& $k=2$ & &  $T_4 D_3, T_5 D_2, T_5 D_3$ \\
& $k=3$ & & $T_4 D_3, T_5 D_2, T_5 D_3$ \\ \hdashline
\multirow{3}{*}{Augmented} & $k=1$ & & $T_2 D_3, T_3 D_2, T_3 D_3, T_4 D_2, T_4 D_3, T_5 D_1, T_5 D_2, T_5 D_3$ \\
& $k=2$ & & $T_2 D_3, T_3 D_2, T_3 D_3, T_4 D_1, T_4 D_2,  T_4 D_3,  T_5 D_1, T_5 D_2, T_5 D_3$ \\
& $k=3$ & & $T_2 D_3, T_3 D_1, T_3 D_2, T_3 D_3, T_4 D_1, T_4 D_2,  T_4 D_3,  T_5 D_1, T_5 D_2, T_5 D_3$\\[6pt]
FDP \\ \cline{1-2}
\multirow{3}{*}{Generalised} & $\gamma = 0.1$ & & $T_2 D_3, T_3 D_2, T_3 D_3,  T_4 D_2, T_4 D_3,  T_5 D_1, T_5 D_2, T_5 D_3$ \\
& $\gamma = 0.2$ & & $T_4 D_3, T_5 D_2, T_5 D_3$\\
& $\gamma = 0.3$ & & $T_4 D_3, T_5 D_2, T_5 D_3$\\ \hdashline
\multirow{3}{*}{Augmented}& $\gamma = 0.1$  & & $T_2 D_3, T_3 D_2, T_3 D_3,  T_4 D_2, T_4 D_3,  T_5 D_1, T_5 D_2, T_5 D_3$\\
& $\gamma = 0.2$  & & $T_2 D_3, T_3 D_1, T_3 D_2, T_3 D_3, T_4 D_1, T_4 D_2, T_4 D_3,  T_5 D_1, T_5 D_2,  T_5 D_3$ \\
& $\gamma = 0.3$  & & $T_2 D_2, T_2 D_3, T_3 D_1, T_3 D_2, T_3 D_3, T_4 D_1, T_4 D_2, T_4 D_3,  T_5 D_1, T_5 D_2,  T_5 D_3$
 \\ [1ex] \hline

\end{tabular}

\singlespacing
\caption{\label{tab:PD_rej1} \textit{Rejected hypotheses for the pharmacodynamic study of~\citet{Ferber2011}, with initial weights of 1/3 for $T_4 D_3, T_5 D_2$ and $T_5 D_3$.}}

\end{table}

There is a similar pattern for the FDP controlling procedures, which is expected given that they are based on the $k$-FWER controlling procedures. For $\gamma =0.1$ the generalised and augmented graphical procedures give the same 8 rejections, which are also the same as the $k$-FWER controlling procedures when $k=1$. For $\gamma = 0.2$ and $\gamma = 0.3$, the augmented procedure rejects 10 and 11 hypotheses respectively. However, again the generalised graphical procedure rejects fewer hypotheses for the larger values of $\gamma = 0.2$ and $\gamma = 0.3$, with only 3 hypotheses rejected. These are the same rejections as the generalised $k$-FWER controlling procedure for $k>1$, because $k_j > 1$ in Algorithm~\ref{alg:FDP}.

We also consider the setting where all 15 hypotheses have initial weight of 1/15. Table \ref{tab:PD_rej2} shows the  resulting rejections for the generalised and augmented graphical $k$-FWER and FDP controlling procedures, with $\delta=1$. 
With these new initial weights, the $k$-FWER controlling procedures both reject the same~7 hypotheses when $k=1$ and the same~8 hypotheses with $k=2$. This shows how the generalised graphical procedure can benefit with non-zero initial weights. However, for $k=3$ the augmented procedure rejects one more hypothesis $(T_5 D_1)$ than the generalised graphical procedure. This is because all hypotheses have fewer than 3 donors, and hence the weights for the generalised graphical procedure cannot increase -- i.e.\ there is no propagation of the significance levels.

Similarly,  the FDP controlling procedures both reject the same 7 hypotheses when $\gamma=0.1$ and the same 8 hypotheses when $\gamma = 0.2$, which are also the same rejections as the $k$-FWER controlling procedures when $k=1$ and $k=2$ respectively. However, for $\gamma = 0.3$ the augmented procedure rejects two more hypotheses than the generalised graphical procedure, while the latter only gives the same rejections as when $\gamma = 0.2$. This is because when $\gamma = 0.3$, $k_j > 2$ in Algorithm~\ref{alg:FDP} and there is no propagation of the significance levels.

\begin{table}[H]
\centering
\doublespacing
\begin{tabular}{l l p{0pt}l }
\\ \hline \Tstrut
\textbf{Procedure} & & & \textbf{Rejected hypotheses} \\ \hline
$k$-FWER \\ \cline{1-2}
\multirow{3}{*}{Generalised} & $k=1$ & & $T_2 D_3, T_3 D_2, T_3 D_3, T_4 D_2, T_4 D_3, T_5 D_2,  T_5 D_3$ \\
& $k=2$ & & $T_2 D_2, T_2 D_3, T_3 D_2, T_3 D_3, T_4 D_2, T_4 D_3, T_5 D_2, T_5 D_3$\\
& $k=3$ & & $T_2 D_2, T_2 D_3, T_3 D_2, T_3 D_3, T_4 D_2, T_4 D_3, T_5 D_2,  T_5 D_3$ \\  \hdashline
\multirow{3}{*}{Augmented}& $k=1$ & & $T_2 D_3, T_3 D_2, T_3 D_3, T_4 D_2, T_4 D_3, T_5 D_2,  T_5 D_3$  \\
& $k=2$ & & $T_2 D_3, T_3 D_2, T_3 D_3, T_4 D_2, T_4 D_3, T_5 D_1, T_5 D_2,  T_5 D_3$\\
& $k=3$ & & $T_2 D_2, T_2 D_3, T_3 D_2, T_3 D_3, T_4 D_2, T_4 D_3, T_5 D_1, T_5 D_2,  T_5 D_3$\\ \vspace{6pt}
FDP \\ \cline{1-2}
\multirow{3}{*}{Generalised} & $\gamma = 0.1$ & & $T_2 D_3, T_3 D_2, T_3 D_3, T_4 D_2, T_4 D_3, T_5 D_2, T_5 D_3$ \\
& $\gamma = 0.2$ & & $T_2 D_2, T_2 D_3, T_3 D_2, T_3 D_3, T_4 D_2, T_4 D_3, T_5 D_2, T_5 D_3$ \\
& $\gamma = 0.3$ & & $T_2 D_2, T_2 D_3, T_3 D_2, T_3 D_3, T_4 D_2, T_4 D_3, T_5 D_2, T_5 D_3$ \\ \hdashline
\multirow{3}{*}{Augmented} & $\gamma = 0.1$ & & $T_2 D_3, T_3 D_2, T_3 D_3, T_4 D_2, T_4 D_3, T_5 D_2, T_5 D_3$ \\
& $\gamma = 0.2$ & & $T_2 D_3, T_3 D_2, T_3 D_3, T_4 D_2, T_4 D_3, T_5 D_1, T_5 D_2, T_5 D_3$\\ & $\gamma = 0.3$ & & $T_2 D_2, T_2 D_3, T_3 D_1, T_3 D_2, T_3 D_3, T_4 D_2, T_4 D_3, T_5 D_1, T_5 D_2, T_5 D_3$
 \\ [1ex] \hline

\end{tabular}

\singlespacing
\caption{\label{tab:PD_rej2} \textit{Rejected hypotheses for the pharmacodynamic study of~\citet{Ferber2011} with initial weights of 1/15 for each hypothesis.}}

\end{table}

\subsection{The Pre-RELAX-AHF trial}
\label{subsec:Pre-RELAX}

Our second case study is a proof-of-concept trial called the Preliminary study of RELAXin in Acute Heart Failure (Pre-RELAX-AHF)~\citep{Teerlink2009}. The trial compared 3 doses of relaxin against a placebo on multiple biological endpoints related to acute heart failure. Given that this was a proof-of-concept trial, less stringent error rates can be used when adjusting for multiplicity.

One criterion for recommending the treatment for further testing is to show an effect on the majority of multiple endpoints. Following~\citet{Davison2011}, we consider a subset of~9 endpoints. We focus on the 30$\mu$g/kg/day dose of relaxin treatment, which showed efficacy on~6 of these endpoints when compared to placebo, using one-sided (uncorrected) $p$-values with $\alpha=0.1$. In what follows, we call the 30$\mu$g/kg/day dose of relaxin treatment the experimental treatment, and the placebo the control treatment.

Since the experimental treatment was declared efficacious in~6 out of~9 endpoints in the pre-RELAX-AHF trial, we consider a trial design where it is required to reject at least~6 out of~9 hypotheses to declare success. Calling these the primary hypotheses, we then add a hierarchical structure to this trial by supposing that we also test secondary hypotheses if at least~6 out of the~9 primary hypotheses were rejected. Hence we have a family of primary hypotheses $\mathcal{F}_1 = (H_1, \ldots, H_9)$ corresponding to testing the experimental treatment against the control across the~9 endpoints, and a family of secondary hypotheses $\mathcal{F}_2$.

We can represent this 6-out-of-9 gatekeeping procedure using entangled graphs, which were described in Section \ref{subsec:entangled}. More precisely, we can define gatekeeping graphs for all ${9 \choose 6} = 84$ possible subsets of~6 primary hypotheses and then entangle them~\citep{Maurer2013}. We perform a Holm procedure $\mathcal{HP}_l$ on~6 hypotheses for each of the~84 subsets of size~6, which we denote $J_l$, $l = 1, \ldots, 84$. The full significance level~$\alpha$ is passed on to $\mathcal{F}_2$ if all 6 hypotheses in $\mathcal{F}_{1l} = \{ H_i : i \in J_l\}$ are rejected. The testing procedure is given by the entangled graph $\mathcal{E} \left( \bm{c},  \mathcal{HP}_l; \,  l = 1, \ldots, 84\right)$ where $c_i = 1/84$ for $i = 1, \ldots, 84$.

This is equivalent to the following testing strategy: the usual Holm procedure is performed on the~9 hypotheses in $\mathcal{F}_1$ at level~$\alpha$ until any six of these hypotheses are rejected. The remaining primary and secondary hypotheses are then tested using the weights given in Table~\ref{tab:gate_weights}, which depend on the number $|I|$ of unrejected hypotheses in $\mathcal{F}_1$. For simplicity, in what follows we suppose that $\mathcal{F}_2$ consists of a single hypothesis $H_{10}$ (which could, for example, represent a composite safety endpoint). We can then use the weights given in Table~\ref{tab:gate_weights} in the $k$-FWER and FDP controlling graphical procedures.

\begin{table}[ht!]
\centering

\begin{tabular}{c p{4pt} c c}
\\ \hline \Tstrut
 $|I|$  & & Weight for each hypothesis in $\mathcal{F}_1$ & Weight for $\mathcal{F}_2 = \{ H_{10} \}$ \\ \hline
$>3$ & & $1/|I|$ & 0 \\
3 & & 83/252 & 1/84 \\
2 & & 11/24 & 1/12 \\
1 & & 2/3 & 1/3 \\
0 & & -- & 1
 \\ [1ex] \hline

\end{tabular}

%\singlespacing
\caption{\label{tab:gate_weights} \textit{Table of weights for the entangled graph procedure used to analyse the trial based on Pre-RELAX-AHF. Here $|I|$ denotes the number of unrejected hypotheses in $\mathcal{F}_1$.}}

\end{table}

In our simulation study, for the primary hypotheses $\mathcal{F}_1$ we follow~\citet{Delorme2016} and take the empirical means and standard errors of the endpoints as the true parameter values for the experimental (E) and control (C) treatments. The numerical values of the means $\mu^C$, $\mu^E$ and standard deviations $\sigma_C$, $\sigma_E$ are given in Appendix~\ref{Asec:pre-relax}. We assume that the distributions of the observed means of the endpoints for the experimental and control treatments follow a multivariate normal distribution: $\bar{X}^G \sim N(\mu^G, \Sigma_G)$ where $G \in \{C, E\}$ and $\Sigma_G =\text{diag}(\sigma_G) \Sigma(\rho) \text{diag}(\sigma_G)$. Here $\text{diag}(\sigma_G)$ is a diagonal matrix with the $i$-th diagonal element equal to $\mu_i^G$, and $\Sigma(\rho)$ is a correlation matrix with ones on the diagonal and $\rho$ on all off-diagonal terms. The test statistic for endpoint $i$ is given by $T_i = \left(\widehat{\text{Var}}(\bar{X}_i^E - \bar{X}_i^C)\right)^{-1/2} \left(\bar{X}_i^E - \bar{X}_i^C \right)$, which is compared with a $t$-distribution. The estimator of the variance of the difference between the means, as well as the appropriate degrees of freedom for the $t$-distribution are given by~\citet{Delorme2016} and implemented in their~R package \texttt{rPowerSampleSize} \citep{deMicheaux2018}.
For the secondary hypothesis $H_{10}$, for simplicity we assume that the test statistic $T_{10}$ follows a normal distribution with mean~3 and variance~1, and is independent of the test statistics for $\mathcal{F}_1$.

Table~\ref{tab:Pre-RELAX} gives the marginal power to reject each hypothesis $H_1, \ldots H_{10}$, calculated using $10^4$ trial replications, with $\alpha=0.1$ and $\delta=1$. The results show that in all scenarios, the augmented procedure has an equal or higher power to reject each of the hypotheses $H_1, \ldots, H_{10}$. For the primary hypotheses $H_1, \ldots, H_5$ and $H_8$, this is especially noticeable for the $k$-FWER controlling procedures when $k = 2$ and $k = 3.$ For hypothesis $H_9$, the augmented procedures have a substantially higher power compared with the generalised graphical procedure (except for when controlling the usual FWER). However, $H_9$ is actually a true null hypothesis (with $\mu_9^C = \mu_9^E = 0.07$) and so this implies a higher type~I error rate for $H_9$ when
using the augmented procedure. In fact the type~I error rate for $H_9$ is below or equal to the nominal 10\% in all scenarios for the generalised graphical procedures. Finally, for the secondary hypothesis $H_{10}$ (which has an initial weight of zero), we see that the power decreases as~$k$ and~$\gamma$ increases for the $k$-FWER and FDP controlling generalised graphical
procedures respectively (in particular, the power is only 6\% when $\gamma = 0.3$ for the latter procedure). Again this shows that in contrast to the augmented procedures, the generalised graphical approaches do not effectively propagate the significance levels when there is a hierarchical structure in the hypotheses.

\begin{table}[ht!]
\centering
\doublespacing
\begin{tabular}{l l p{0pt} c c c c c c c c c c }
\\ \hline \Tstrut
\textbf{Procedure} & & & \multicolumn{10}{c}{\textbf{Marginal power}} \\ \hline
& & & $H_1$ & $H_2$ & $H_3$ & $H_4$ & $H_5$ & $H_6$ & $H_7$ & $H_8$ & $H_9$ & $H_{10}$ \\[-6pt]
$k$-FWER \\ \cline{1-2}
\multirow{2}{*}{$k = 1$} & Generalised & &  95 & 89 & 72 & 78 &  85 & 100 & 100 & 62 &   6 & 64\\
& Augmented & & 95 & 89 & 72 & 78 &  85 & 100 & 100 & 62 &   6 & 64 \\ \hdashline
\multirow{2}{*}{$k = 2$} & Generalised & &  97 &  93 &  79 &  84 &  90 &  100 & 100 &  70 &  9 &  60 \\
& Augmented & & 99 &  97 &  90 & 92 & 95 & 100 & 100 & 87 & 65 & 84\\ \hdashline
\multirow{2}{*}{$k = 3$} & Generalised & & 98 & 95 & 81 & 86 & 92 & 100 & 100 & 73 &    10 & 42 \\
& Augmented & & 100 & 99 & 96 & 97 & 98 & 100 & 100 & 95 & 87 & 95\\[6pt]
FDP \\ \cline{1-2}
\multirow{2}{*}{$\gamma = 0.1$} & Generalised & & 95 & 89 & 72 & 78 & 85 & 100 & 100 &  62 & 8 & 61\\
& Augmented & &  95 & 89 & 72 & 78 & 85 & 100 & 100 & 63 & 38 & 65 \\ \hdashline
\multirow{2}{*}{$\gamma = 0.2$} & Generalised & & 95 & 92 & 78 & 83 & 89 & 100 & 100 & 70 & 9 & 50 \\
& Augmented & &  96 & 93 & 82 & 86 & 90 & 100 & 100 & 77 & 52 & 73\\ \hdashline
\multirow{2}{*}{$\gamma = 0.3$} & Generalised & & 96 & 93 & 80 & 85 & 90 & 100 & 100 & 72 & 10 & 6 \\
& Augmented & & 97 & 94 & 86 &  88 & 92 & 100 & 100 & 83 & 58 & 83
 \\ [1ex] \hline

\end{tabular}
\singlespacing
\caption{\label{tab:Pre-RELAX} \textit{Simulated marginal powers to reject hypotheses $H_1, \ldots, H_{10}$, with $\alpha = 0.1$ and the distribution of the test statistics for $\mathcal{F}_1 = (H_1, \ldots, H_9)$ based on the Pre-RELAX-AHF trial reported by \citet{Teerlink2009}. Results are based on $10^4$ independent trial replications.}}

\end{table}

\subsection{ATMOSPHERE study}
\label{subsec:atmosphere}

Our final case study is motivated by the confirmatory ATMOSPHERE study~\citep{Krum2011} in patients with heart failure. As described in~\citet{Maurer2014}, the trial compared three therapies: aliskiren monotherapy~(A), enalapril monotherapy~(E), and aliskiren/enalapril combination therapy~(C). This resulted in three single primary hypotheses ($H_1, H_2, H_3$) and two families of secondary hypotheses ($\mathcal{H}_4, \mathcal{H}_5$):

\begin{itemize}
\singlespacing

\item[] $H_1$: non-superiority of C versus E

\item[] $H_2$: inferiority of A versus E

\item[] $H_3$: non-superiority of A versus E

\item[] $\mathcal{H}_4 = \{ H_{41}, H_{42} \}$: two secondary endpoints for comparing C versus E

\item[] $\mathcal{H}_5 = \{ H_{51}, H_{52} \}$: two secondary endpoints for comparing A versus E \vspace{-6pt}

\end{itemize}

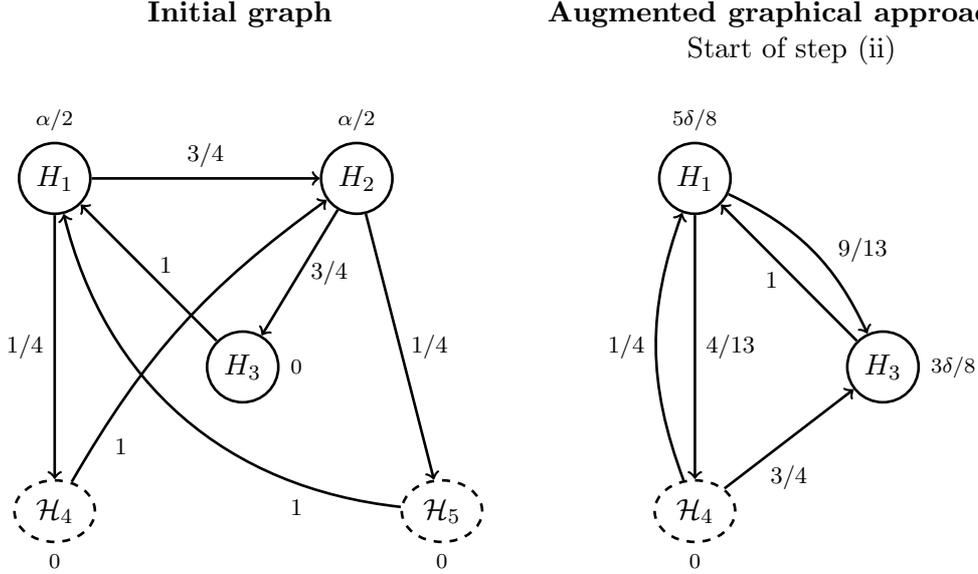
\begin{figure}[ht!]
\centering
\begin{tabular}{c p{12pt} c}

\textbf{Initial graph} & & \textbf{Augmented graphical approaches} \\ & & Start of step (ii) \\[12pt]

\begin{tikzpicture}[line width = 1]
% nodes %
\node[circle, draw, text centered, label=above:{\scriptsize $\alpha/2$}] (H1) {$H_1$};
\node[circle, draw, right = 3 of H1, text centered, label=above:{\scriptsize $\alpha/2$}] (H2) {$H_2$};
\node[circle, draw, right = 1.5 of H1, yshift=-2.5cm, text centered, label=right:{\scriptsize $0$}] (H3) {$H_3$};
\node[ellipse, draw, dashed, below=3.5 of H1, text centered, label=below:{\scriptsize $0$}] (H4) {$\mathcal{H}_4$};
\node[ellipse, draw, dashed, below=3.5 of H2, right=4 of H4,  text centered, label=below:{\scriptsize $0$}] (H5) {$\mathcal{H}_5$};
 
% edges %
\draw[->] (H1)-- node[above,font=\footnotesize]{$3/4$}  (H2);
\draw [->] (H1) -- node[left,font=\footnotesize]{$1/4$}  (H4);
\draw [->] (H3) -- node[right, yshift = 0.1cm, font=\footnotesize]{$1$}  (H1);
\draw [->] (H2) -- node[right,font=\footnotesize]{$3/4$}  (H3);
\draw [->] (H2) -- node[right,font=\footnotesize]{$1/4$}  (H5);
\path [->] (H4) edge [bend left=12] node[yshift = -1.6cm, xshift = -0.8cm,font=\footnotesize]{$1$}  (H2);
\path [->] (H5) edge [bend left=35] node[yshift = -1.2cm, xshift = 1.5cm,font=\footnotesize]{$1$}  (H1);

\end{tikzpicture}
& &
\begin{tikzpicture}[line width = 1]
% nodes %
\node[circle, draw, text centered, label=above:{\scriptsize $5\delta/8$}] (H1) {$H_1$};
\node[circle, draw, right = 1.5 of H1, yshift=-2.5cm, text centered, label=right:{\scriptsize $3\delta/8$}] (H3) {$H_3$};
\node[ellipse, draw, dashed, below=3.5 of H1, text centered, label=below:{\scriptsize $0$}] (H4) {$\mathcal{H}_4$};

% edges %
\path [->] (H1) edge [bend left=0] node[right,font=\footnotesize]{$4/13$}  (H4);
\path [->] (H1) edge [bend left=20] node[right,xshift = 0.2cm, font=\footnotesize]{$9/13$}  (H3);
\path [->] (H3) edge [bend left=0] node[left, yshift = -0.1cm, font=\footnotesize]{$1$}  (H1);
\path [->] (H4) edge [bend left=20] node[left,font=\footnotesize]{$1/4$}  (H1);
\draw[->] (H4)-- node[below, yshift = -0.2cm, font=\footnotesize]{$3/4$}  (H3);

\end{tikzpicture}

\\[12pt]
%
%\textbf{After rejecting $H_2$} &&  \textbf{Final graph} \\[12pt]
%
%%
%& & 
%%

\end{tabular}

\caption{The graph on the left-hand side was used for the ATMOSPHERE study, as presented in~\citet{Maurer2014}. The graph on the right-hand side is the updated graph at the start of step~(ii) in the augmented graphical approach for either $k$-FWER or FDP control, after $H_2, H_{51}$ and $H_{52}$ have been rejected.}

\label{fig:ATMOSPHERE}

\end{figure}

\noindent The graph on the left-hand side in Figure~\ref{fig:ATMOSPHERE} shows the graphical test procedure used in~\citet{Maurer2014} to analyse the trial. Note that if all individual null hypotheses in $\mathcal{H}_4$ or $\mathcal{H}_5$ are rejected, the local significance level is propagated to the remaining hypotheses. For simplicity, we apply a Holm procedure within each of the two secondary families $\mathcal{H}_4$ and $\mathcal{H}_5$. Following \cite{Maurer2014}, suppose we observe the (hypothetical) unadjusted $p$-values $p_1 = 0.1$, $p_2 = 0.007$, $p_3 = 0.05$, $p_{41} = 0.0015$, $p_{42} = 0.04$, $p_{51} = 0.0031$ and $p_{52} = 0.001$.  \\

Consider first controlling the $k$-FWER with $k=2$ and $\alpha=0.025$. For the augmented graphical approach (given in Algorithm~\ref{alg:kFWER_aug}), in step~(i) the Bonferroni-based graphical procedure for FWER control would reject $H_2$, $H_{51}$ and $H_{52}$. The updated graph used at the start of step~(ii) is shown in the right-hand side of Figure~\ref{fig:ATMOSPHERE}, where $\alpha$ has been replaced by $\delta$. Supposing that $\delta = 0.5$, step~(ii) of the algorithm rejects $H_3$. Since we have made one additional (augmented) rejection, at this point we stop.  As for the generalised graphical approach for $k$-FWER control (given in Algorithm~\ref{alg:kFWER}), in step~(ii) we would only reject $H_2$. Since the number of rejections $|R| = k-1$, we stop at this point.

Now consider controlling the FDP with $\gamma = 0.3$. For the augmented graphical approach (given in Algorithm~\ref{alg:FDP_aug}), in step~(i) we reject $H_2$, $H_{51}$ and $H_{52}$ like before. In step~(ii), we can reject one additional hypothesis, and hence we reject $H_3$ and then stop. Finally, for the generalised graphical approach (given in Algorithm~\ref{alg:FDP}), we first apply the usual Bonferroni-based graphical procedure for FWER control, which rejects $H_2$, $H_{51}$ and $H_{52}$. Since $|R_1| > 1/\gamma -1$, we then apply the 2-FWER procedure which (as above) only rejects $H_2$. Since $|R_2| < 2/\gamma -1$, we stop and only reject $H_2$.

%%%%%%%%%%%%%%%%%%%%%%%%%%%%%%%%%%%%%%%%%%%%%%%%%%%%
\section{Discussion}
\label{sec:discuss}

In this paper, we have showed how to generalise the graphical approach of hypothesis testing \citep{Bretz2009} so that the $k$-FWER or the FDP can be controlled. By applying the methodology of~\citet{Romano2010} and~\citet{Laan2004}, we have proposed generalised and augmented graphical approaches for both $k$-FWER and FDP control (as well as an augmented procedure for asymptotic FDR control). Crucially, these approaches respect the hierarchical structure of the underlying multiple testing procedure given by the graphical weighting strategy. We have also applied the proposed graphical approaches to three real-life case studies covering a broad range of clinical trial applications.

Our recommendation is that
%in the context of clinical trials, where typically the number of hypotheses is relatively small ($<100$),
the augmented graphical approaches should be used instead of the generalised graphical approaches. Firstly, the generalised graphical approach for $k$-FWER control has the undesirable property that if a hypothesis~$H_j$ has fewer than~$k$ donors, its initial significance level will not increase. Hence, the generalised graphical approach cannot effectively propagate the significance levels through the graph. The case studies in Section~\ref{sec:case_study} show how this can have a detrimental effect on the power of the generalised graphical approach -- the power to reject hypotheses with fewer than~$k$ donors can actually decrease as~$k$ increases. Since the generalised graphical approach for FDP control is based on the generalised graphical approach for $k$-FWER control, a similar problem occurs. %-- the power to reject hypotheses with fewer than~$k$ donors can decrease as~$\gamma$ increases.

In contrast, the augmented graphical approach is able to propagate significance levels to all hypotheses that have fewer than~$k$ donors. As a consequence, the power of the augmented graphical approach for $k$-FWER control and FDP control increases as~$k$ and~$\gamma$ increase (respectively). Importantly, in all of the case studies in Section~\ref{sec:case_study}, the augmented graphical approach had a higher power (or rejected at least as many hypotheses) compared with the generalised graphical approach. These results are backed up by existing power comparisons for the generalised and augmented Holm procedure \citep{Romano2007, Dudoit2004} when testing a relatively small number of hypotheses.

The research for this paper was motivated by clinical trial applications ranging from early to late drug development, as illustrated by the case studies in Section~\ref{sec:case_study}. Outside of the context of clinical trials and the graphical weighting strategy of~\citet{Bretz2009}, another area of application is testing hypotheses in a directed acyclic graph (DAG) for use in gene set analysis, as proposed by~\citet{Meijer2015}. The authors presented a top-down method that strongly controls the FWER, and by considering the genes and gene sets as nodes in a DAG, the method allows testing for simultaneous testing of both significant gene sets and individual genes. The testing procedure starts with an initial weight for each of the leaf nodes (i.e.\ nodes without any descendants), and an iterative weighting procedure is used to update the weights for all the other nodes in the graph. These weights also satisfy the monotonicity condition given in equation~\eqref{eq:monotonicity}, and so  suitably modified versions of the augmented and generalised graphical approaches could be used in this setting.

As future work, it would be desirable to derive adjusted $p$-values for all of the proposed procedures, especially for the augmented graphical approaches. This would involve extending the results of ~\citet{Laan2004}, who showed how to calculate adjusted $p$-values for their augmented approach. Finally, the initial motivation for this paper came from considering the generalised closure principle~\citep{Guo2010}, which was applied to derive stepup procedures for $k$-FWER control. The usual graphical approach for FWER corresponds to defining a shortcut closed testing procedure~\citep{Bretz2009}. It would be interesting to formalise a similar link between the generalised graphical approach for $k$-FWER control and the generalised closure principle.

%%%%%%%%%%%%%%%%%%%%%%%%%%%%%%%%%%%%%%%%%%%%%%%%%%%

\section*{Acknowledgements}

The authors would like to thank Olivier Guilbaud for suggesting the use of adjusted $p$-values for the augmented approach for $k$-FWER control. This work was funded by the UK Medical Research Council (MC\_UU\_00002/6 (JMSW and DSR) and  MR/N028171/1 (JMSW)) and the Biometrika Trust (DSR).

\section*{Data availability statement}

All of the data that support the findings of this study are available within the article itself. Code to reproduce the results of Section~\ref{sec:case_study} can be found at \url{https://github.com/dsrobertson/graphical-approach}

%%%%%%%%%%%%%%%%%%%%%%%%%%%%%%%%%%%%%%%%%%%%%%%%%%%

%\clearpage

{%\singlespacing
\bibliography{graphical_generalised_error}
}

%%%%%%%%%%%%%%%%%%%%%%%%%%%%%%%%%%%%%%%%%%%%%%%%%%%

%\newpage

\appendix

% Algorithms
\setcounter{algorithm}{0}
\renewcommand{\thealgorithm}{\Alph{section}\arabic{algorithm}}
\counterwithin*{algorithm}{section}

%\singlespacing
\section{Graphical weighting strategy and weighted Bonferroni test for FWER control}
\label{Asec:graphFWER}
%\doublespacing

For a given index set $J \subseteq M$, let $J^c = M \! \setminus \! J$ denote the set of indices not contained in~$J$

\begin{algorithm}[ -- Graphical weighting strategy] \hspace{-6pt} \citep[Algorithm 1]{Bretz2011} \label{alg:graph_weight} 
\hspace{0pt} \\[-24pt]
\begin{enumerate}[label=(\roman*)]
\singlespacing
\item Set $I = M$
\item Select $j \in J^c$ and remove $H_j$
\item Update the graph: \begin{align*}
I & \rightarrow I\setminus \! \{j\}, J^c \rightarrow J^c \setminus \! \{ j \} \\
w_l(I) & \rightarrow \begin{cases}
w_l(I) + w_j(I) g_{jl} & l \in I \\
0 & \text{otherwise}
\end{cases} \\
g_{lh} & \rightarrow \begin{cases}
\displaystyle\frac{g_{lh} + g_{lj}g_{jh}}{1 - g_{lj}g_{jl}} & l,h \in I, \, l \neq h, \, g_{lj}g_{jl} < 1 \\
0 & \text{otherwise}
\end{cases}
\end{align*}
\item If $|J^c| \geq 1$, go to step (ii); otherwise set $w_l(J) = w_l(I), l \in J$ and stop.
\end{enumerate}

\end{algorithm}

\noindent The weights $w_j(J), j \in J$, generated by this procedure are unique~\citep{Bretz2009}, and in particular do not depend on which order the hypotheses $H_j, j \in J^c$ are removed in Algorithm~\ref{alg:graph_weight}.

\newpage

\begin{algorithm}[ -- Bonferroni-based graphical test for FWER control] \hspace{-6pt} \citep[Algorithm 2]{Bretz2011}  \label{alg:FWER} 
\hspace{0pt} \\[-24pt]
\begin{enumerate}[label=(\roman*)]
\singlespacing
\item Set $I = M$
\item Select a $j \in I$ such that $p_j \leq w_j(I)\alpha$ and reject $H_j$; otherwise stop.
\item Update the graph: \begin{align*}
I & \rightarrow I\setminus \{j\} \\
w_l(I) & \rightarrow \begin{cases}
w_l(I) + w_j(I) g_{jl} & l \in I \\
0 & \text{otherwise}
\end{cases} \\
g_{lh} & \rightarrow \begin{cases}
\displaystyle\frac{g_{lh} + g_{lj}g_{jh}}{1 - g_{lj}g_{jl}} & l,h \in I, \, l \neq h, \, g_{lj}g_{jl} < 1 \\
0 & \text{otherwise}
\end{cases}
\end{align*}
\item If $|I| \geq 1$, go to step (ii); otherwise stop.
\end{enumerate}
\end{algorithm}

\noindent The final decisions of the algorithm do not depend on which order the hypotheses are rejected. For example, step~(ii) above could be replaced by choosing $j = \arg \min_{i \in I}\{p_i/w_i(I)\}$. \\

\begin{algorithm}[ -- Adjusted $p$-values] \hspace{-6pt} \citep[Algorithm 2]{Bretz2009} \label{alg:adjp} 
\hspace{0pt} \\[-24pt]
\begin{enumerate}[label=(\roman*)]
\singlespacing
\item Set $I = M$ and $p_{\max} = 0$
\item Let $j = \arg \min_{i \in I} p_i / w_i(I)$
\item Calculate $p_j^{\text{adj}} = \max \{ p_j/w_j(I), p_{\max} \}$ and set $p_{\max} = p_j^{\text{adj}}$.
\item Update the graph: \begin{align*}
I & \rightarrow I\setminus \{j\} \\
w_l(I) & \rightarrow \begin{cases}
w_l(I) + w_j(I) g_{jl} & l \in I \\
0 & \text{otherwise}
\end{cases} \\
g_{lh} & \rightarrow \begin{cases}
\displaystyle\frac{g_{lh} + g_{lj}g_{jh}}{1 - g_{lj}g_{jl}} & l,h \in I, \, l \neq h, \, g_{lj}g_{jl} < 1 \\
0 & \text{otherwise}
\end{cases}
\end{align*}
\item If $|I| \geq 1$, go to step (ii); otherwise stop.
\item Reject all hypotheses $H_j$ with $p_j^{\text{adj}} \leq \alpha$
\end{enumerate}
\end{algorithm}

\newpage

%%%%%%%%%%%%%%%%%%%%%%%%%%%%%%%%%%%%%%%%%%%%%%%%%%%
\section{Further results on the generalised graphical approach}

%\singlespacing
\subsection{Streamlined and operative versions of the generalised graphical approach}
\label{Asubsec:streamlined}

%\doublespacing

Apart from special cases, applying the generalised graphical approach given in Algorithm~\ref{alg:kFWER} can be computationally intensive for larger values of~$m$, particularly since the weights $w_i(K)$ need to be calculated using Algorithm~\ref{alg:graph_weight}. For large values of~$m$, we can directly apply the streamlined version of the general stepdown method for controlling the $k$-FWER in~\citet[Algorithm 4.2]{Romano2010} to give a streamlined version of Algorithm~\ref{alg:kFWER}. The way this version works is to avoid minimising over all subsets of size $k-1$ of previously rejected hypothesis, and only consider the least significant $k-1$ of the previous rejections. Note that this only gives asymptotic control of the $k$-FWER (as the sample size of the trial increases).

\begin{algorithm}[ -- Streamlined graphical approach for $k$-FWER control] \label{alg:kFWER_streamlined}
\hspace{0pt} \\[-24pt]

\noindent Given an index set~$R$ of rejected hypohteses, let $p_{1:R} \leq p_{2:R} \leq \cdots \leq  p_{|R|:R}$ denote the ordered $p$-values, with corresponding hypotheses $H_{1:R}, H_{2:R}, \ldots, H_{|R|:R}$. Denote by $\{r_1, \ldots, r_{|R|}\}$ the permutation of $\{1, \ldots, |R|\}$ that gives this ordering, so that $p_{1:R} = p_{r_1}, \ldots, p_{|R|:R} = p_{r_{|R|}}$.
The streamlined algorithm is the same as Algorithm~\ref{alg:kFWER}, except that at step~(iv) we now reject any $H_i$, $i \in I$ for which $p_i \leq w_i(K) k \alpha$, where $
K = I \cup \{r_{|R|-k+2}, \ldots, r_{|R|} \}$.
\end{algorithm}

The streamlined version only gives asymptotic control of the $k$-FWER, but involves no minimisation over any subsets. In order to get closer to the original, exact algorithm while still retaining computational feasibility, as a compromise we can use the operative method proposed in~\citet[Remark 3.3]{Romano2010}. Consider that to compute the critical value in step~(iv) of Algorithm~\ref{alg:kFWER}, one has to evaluate ${|R| \choose k-1}$ weights in order to choose the minimum. The operative method maximises over subsets not necessarily of the entire index set~$R$ of previously rejected hypotheses, but only for some number~$B$ least significant hypotheses so far. More precisely, we have the following algorithm:

\begin{algorithm}[ -- Operative graphical approach for $k$-FWER control] \label{alg:kFWER_operative}
\noindent Pick a user-specified number $N_{\max}$ and let $B$ be the largest integer for which ${B \choose k-1} \leq N_{\max}$. The operative method is the same as Algorithm~\ref{alg:kFWER}, except that step~(iv) rejects any $H_i$, $i \in I$ for which  \[
p_i \leq \min_{J \subseteq \{r_{\max\{1, |R|-B+1\}}, \ldots, r_{|R|}\},|J| = k-1} \{w_i(K) : K = I \cup J \} k \alpha
\]
\end{algorithm}

\noindent When $B \geq |R|$ we maximise over all subsets of $R$ of size $k-1$ like in the original algorithm, while the streamlined algorithm is a special case of the operative method where $N_{\max} = 1$ and hence $B = k-1$.

%%%%%%%%%%%%%%%%%%%%%%%%%%%%%%%%%%%%%%%%%%%%%%%%%%%

%\singlespacing
\subsection{Examples of the generalised graphical approach}
\label{Asubsec:example_kgMCP}
%\doublespacing

 % In what follows,  we set $\alpha^* = 0$ for the sub-procedure in step~(iii) of Algorithm~\ref{alg:kFWER} if  $|R| < k-1$.

\begin{example}[ -- Generalised Weighted Bonferroni]

\noindent Suppose each vertex on the graph is unconnected, i.e.\ $g_{ij} = 0$ for all $i, j \in M$. Algorithm~\ref{alg:graph_weight} implies that $w_j(J) = w_j(M)$, $j \in J$ for all $J \subseteq M$. Hence the inequality in step~(iv) of Algorithm~\ref{alg:kFWER} is simply $p_i \leq w_i(M)k\alpha$ and so there is no further testing after step~(ii), unless $\delta >0$ and $|R| < k-1$. Thus when $\delta =0$, Algorithm~\ref{alg:kFWER} is exactly the same as the generalised weighted Bonferroni procedure in Section~\ref{sec:framework} with $w_i = w_i(M)$.
\end{example}

\begin{example}[ -- Generalised Holm]

\noindent To represent the Holm procedure with~$m$ hypotheses, we set the initial weights $w_i(M) = 1/m$ and $g_{ij} = 1/(m-1)$ for all $i,j \in M$, $i \neq j$. Hence using Algorithm~\ref{alg:graph_weight}, we have $w_i(I) = 1/|I|$ for all $i \in I$ and $I \subseteq M$, and the inequality in step~(iv) of Algorithm~\ref{alg:kFWER} is simply \[p_i \leq \frac{k \alpha}{|I| + k - 1} = \frac{k \alpha} { m +k - |R| - 1}\] For $\delta =0$, this gives identical rejections to the generalised Holm procedure as given in~\citet{Lehmann2005}.

\end{example}

%\begin{example}(\textit{$m = 4$})
%
%\noindent Suppose $m = 4$ and $k =2$. Then an example realisation of Algorithm~\ref{alg:kFWER} is as follows:
%
%\begin{enumerate}%[label=(\roman*)]
%\singlespacing
%\item Set $I = \{1,2,3,4\}$.
%%
%\item We reject any $H_i$, $i \in I$ for which $p_i \leq w_i(I)k\alpha$. Suppose $H_1$ and $H_2$ are rejected at this step, hence $R = \{1, 2\}$ and $I = \{3, 4 \}$.
%%
%\item We reject $H_i$, $i \in I$ if $p_i \leq \min \{ w_i(\{1,3,4\}), w_i(\{2,3,4\}) \}$. Suppose only $H_3$ is rejected at this step, hence $R = \{1,2,3\}$ and $I = \{ 4\}$.
%%
%\item We reject $H_4$ if $p_4 \leq \min \{ w_4(\{1,4\}), w_4(\{2,4\}), w_4(\{3,4\}) \}$ and stop.
%\end{enumerate}
%\end{example}

\begin{example}[ -- Hierarchical testing: fixed sequence test and fallback procedure]

\noindent In a fixed sequence test, the hypotheses are tested in a pre-specified order. This allows each hypothesis to be tested at the full level~$\alpha$ while controlling the FWER, with the proviso that if any hypothesis is not rejected then no further testing is allowed. Suppose the pre-specified ordering for testing~$m$ hypotheses is $H_1 \rightarrow H_2 \rightarrow \cdots \rightarrow H_m$. Hence we have $g_{ij} = 1$ for $i = 1, \ldots, m-1$ if $j = i+1$ and $g_{ij} = 0$ otherwise. 

If we follow the usual fixed sequence test and set $w_1(M) = 1$ and $w_i(M) = 0$, $i = 1, \ldots, m-1$, then only $H_1$ can be rejected in step~(ii) of the Algorithm~\ref{alg:kFWER} and hence the algorithm will never proceed to step~(iv) since $|R| < k$. Hence, a more natural generalisation of the fixed sequence test is to set the initial weights as $w_i(M) = 1/k$ for $i = 1, \ldots, k$ and $w_i(M) = 0$ otherwise. This means that the first~$k$ hypotheses will be tested at full level~$\alpha$. However, assume that the first~$k$ hypotheses are all rejected (otherwise we proceed to the sub-procedure of step~(iii) and can only reject up to the first~$k-1$ hypotheses). Since $w_i(\{1, \ldots, k-1, k+1, \ldots, i, \ldots, m\}) = 0$ for $i = k+1, \ldots, m$, step~(iv) of the algorithm implies that no further hypotheses can then be rejected. For $k > 1$, this generalisation of the fixed sequence test has the undesirable property that only the first~$k$ hypotheses can ever be tested, even when using the sub-procedure of step~(iii) with $\delta > 0$.

A similar issue occurs when generalising the fallback procedure~\citep{Wiens2003}, which is a modification of the fixed sequence procedure where the initial weights $w_i(M) > 0$ for all $i \in M$. Applying Algorithm~\ref{alg:kFWER}, suppose (without loss of generality, by relabelling the hypothesis labels) that the hypotheses $H_1, \ldots, H_k$ are all rejected at step~(ii). However, since $w_i(I \bigcup \{1, \ldots, k-1\} \}) = w_i(M)$ for all $i \in I$ and $I \subseteq \{k+1, \ldots, m \}$, step~(iv) implies that the hypotheses $H_i$, $i = k+1, \ldots, m$, are also tested at significance level $w_i(M)k\alpha$. So for $k > 1$, this generalisation of the fallback procedure has the undesirable property that rejecting hypotheses does not lead to an increase in the significance levels of the remaining hypotheses, except via the sub-procedure of step~(iii) when $\delta > 0$ (but even then, the propagation is limited to at most $k-1$ hypotheses).

\end{example}

\begin{example}[ -- Hypotheses with fewer than $k$ donors]

\noindent We can generalise the previous example to any graph where any hypothesis has fewer than~$k$ donors, where the donors of a hypothesis~$H_j$ are the hypotheses that donate (or propagate) their significance levels to $H_j$ if they are rejected. More formally, we denote the donors of hypothesis~$H_j$ by $\text{do}(H_j) = \{H_i : g_{ij}>0\}$. Note that two hypotheses~$H_i$ and $H_j$ can be donors to each other.

If a hypothesis~$H_j$ in a graph has fewer than~$k$ donors, then applying the generalised graphical approach has the undesirable property that the initial significance level for~$H_j$ can never increase, even if all its donors are rejected (except for up to $k-2$ hypotheses via the sub-procedure in step~(iii)). To see this, suppose $\delta = 0$ and all donors of~$H_j$ have been rejected (and $|R| \geq k$, or else there is no propagation). In step~(iv) of Algorithm~\ref{alg:kFWER}, since $|\text{do}(H_j)| \leq k-1$ then \[ \min_{J \subseteq R, |J| = k-1} \{w_i(K) : K = I \cup J \} k\alpha \leq w_i \left(I \cup \text{do}(H_j) \right) k\alpha = w_i(M) k \alpha\] Hence $H_j$ is tested using the initial weights in step~(iv). In particular, this means that if a hypothesis with fewer than~$k$ donors has an initial weight of zero, then it can never be rejected (except possibly via the sub-procedure in step~(iii)). This can be an undesirable property to have in a testing procedure which has a hierarchical structure, as we will see further in the case studies in Section~\ref{sec:case_study}.

As an example, consider the graph for the diabetes trial shown in Figure~\ref{fig:diabetes_trial}, and suppose we wish to control the $k$-FWER for $k=2$. Since the secondary hypotheses~$H_3$ and~$H_4$ only have one donor each (hypotheses~$H_1$ and~$H_2$, respectively) and start with a weight of zero,  they will never be rejected even if more than one of the primary hypotheses~$H_1$ and~$H_2$ are rejected. % We give a numerical demonstration of this in Appendix~\ref{Asec:diabetes_trial}.

% Using the sub-procedure in step~(iii) does not help in this case, since if $|R| < k-1 =1$ then only $H_1$ or $H_2$ can be rejected since they are the only hypotheses with non-zero weights.

\end{example}

\section{Parameter values for the Pre-RELAX-AHF trial}
\label{Asec:pre-relax}

%\doublespacing

In our simulation study, for the primary hypotheses $H_1, \ldots, H_9$ we take the empirical means and standard errors of the endpoints as the true parameter values for the experimental~(E) and control~(C) treatments. The numerical values of the means $\mu^C$, $\mu^E$ and standard deviations $\sigma_C$, $\sigma_E$ are given in~\citet{Delorme2016} and reproduced below: \begin{align*}
\mu^C & = (0.23, 1679, 0.79, -12, 44.2, 0.828, 0.857, 0.13, 0.07) \\
\mu^E & = (0.4, 2567, 0.88, -10.2, 47.9, 0.974, 1, 0.21, 0.07) \\[6pt]
\sigma_C & = (\sqrt{0.23(1-0.23)}, 2556, \sqrt{0.79(1-0.79)}, 7.3, 14.2, \sqrt{0.828(1-0.828)}, \sqrt{0.857(1-0.857)}, \\ & \qquad \sqrt{0.13(1-0.13)}, \sqrt{0.07(1-0.07)}) \\
& = (0.421, 2556, 0.407, 7.3, 14.2, 0.377, 0.350, 0.336, 0.255) \\[6pt]
\sigma_E & = (\sqrt{0.4(1-0.4)}, 2898, \sqrt{0.88(1-0.88)}, 6.1, 10.1, \sqrt{0.974(1-0.974)}, 10^{-12}, \sqrt{0.21(1-0.21)}, \\ & \qquad \sqrt{0.07(1-0.07)}) \\
& =  (0.490, 2898, 0.325, 6.1, 10.1, 0.159, 10^{-12}, 0.407, 0.255)
\end{align*}

\end{document}